\documentclass[floats,showpacs,aps,twocolumn,prb]{revtex4}
\usepackage{graphicx,graphics,epsfig}
\usepackage{dcolumn}
\usepackage{bm,amsmath,amssymb}

\begin{document}

\title{Anderson localization of classical waves in weakly scattering
metamaterials}
\author{Ara A. Asatryan$^1$, Sergey A. Gredeskul$^{2,3}$, Lindsay C. Botten$%
^1$, Michael A. Byrne$^1$, Valentin D. Freilikher$^{4}$, Ilya V. Shadrivov$%
^3 $, Ross C. McPhedran$^5$, and Yuri S. Kivshar$^3$}
\affiliation{$^1$Department of Mathematical Sciences and Centre for Ultrahigh-bandwidth
Devices for Optical Systems (CUDOS), University of Technology, Sydney, NSW
2007, Australia\\
$^2$ Department of Physics, Ben Gurion University of the Negev, Beer Sheva,
84105, Israel\\
$^3$ Nonlinear Physics Center and CUDOS, Australian National University,
Canberra, ACT 0200, Australia\\
$^4$ Department of Physics, Bar-Ilan University, Raman-Gan, 52900, Israel\\
$^5$School of Physics and CUDOS, University of Sydney, Sydney, NSW 2006,
Australia}

\begin{abstract}
We study the propagation and localization of classical waves in one-dimensional disordered structures composed of alternating layers of left- and right-handed materials (mixed stacks) and compare them with structures composed of different layers of the same material (homogeneous stacks).  For weakly scattering layers, we have developed an effective analytical approach and have calculated the transmission length within a wide range of the input parameters. This enables us to describe, in a unified way, the localized and ballistic regimes as well as the crossover between them. When both refractive index and layer thickness of a mixed stack are random, the transmission length in the long-wave range of the localized regime exhibits a quadratic power wavelength dependence with  different coefficients of proportionality  for mixed and homogeneous stacks. Moreover, the transmission length of a mixed stack differs from the reciprocal of the Lyapunov exponent of the corresponding infinite stack. In both the ballistic regime of a mixed stack and in the near long-wave region of a homogeneous stack, the transmission length of a  realization is a strongly fluctuating quantity. In the far long-wave part of the ballistic region, the homogeneous stack  becomes effectively uniform and the transmission length fluctuations are weaker. The crossover region from the localization to the ballistic regime is relatively narrow for both mixed and homogeneous stacks. In mixed stacks with only refractive-index disorder, Anderson localization at long wavelengths is substantially suppressed, with the localization length growing with  wavelength much faster than for homogeneous stacks. The crossover region becomes essentially wider and transmission resonances appear only in much longer stacks. The effects of absorption on  one-dimensional transport and localization have also been studied, both analytically and numerically. Specifically, it is shown that the crossover region is particularly sensitive to losses, so that even small absorption noticeably suppresses frequency dependent oscillations in the transmission length. All theoretical predictions are in an excellent agreement with the results of numerical simulations.
\end{abstract}

\maketitle

\bigskip

%%%%%%%%%%%%%%%%%%%%%%%%%%%%%%%%%%%%%%%%%%%%%%%%%%

\section{Introduction}

\label{sec:intro} %%%%%%%%%%%%%%%%%%%%%%%%%%%%%%%%%%%%%%%%%%%%%%%%%%%%%%%

Metamaterials are artificial structures having negative refractive indices for some wavelengths~\cite{Veselago}.  While natural materials having such properties are not known, it was the initial paper \cite{Pendry} that sought to  realize artificial metamaterials which triggered the rapidly increasing interest in this topic.  Over the past decade, the physical properties of these structures, and their possible applications in modern optics and microelectronics, have received considerable attention (see e.g. Refs  \cite{Sok,Bliokh,classification,Shalaev}). The reasons for such interest are their unique physical properties, their ability to overcome the diffraction limit~\cite{Veselago,Pendry}, and their potential role in cloaking~\cite{Schurig}, the suppression of spontaneous emission rate \cite{emis}, and the enhancement of  quantum interference \cite{inter}, etc.

Until recently,  most studies considered only ideal systems and did not address the possible effects of disorder. However, real metamaterials are always disordered, at least, in part, due to fabrication errors.  Accordingly, the study of disordered metamaterials is not just an academic question but is also relevant to their application.   The first step in this direction was made in Ref.~\cite{defect} where it was shown that the presence of a single defect led to the appearance of a localized mode. A metamaterial with many point-like defects was considered in Ref.~\cite{Gorkunov} where it was demonstrated that even weak microscopic disorder might lead to a substantial suppression of wave propagation through a metamaterial over a wide frequency range.

The next steps in this direction focused on the study of  localization in metamaterials. Anderson localization is one of the most fundamental and fascinating phenomenon of the physics of disorder. Predicted in the seminal paper~\cite{And} for spin excitations, it was extended to the case of electrons and other one-particle excitations in solids, as well as to electromagnetic waves (see, e.g., Refs.~\cite{john84,LGP,Ping91,FG,GMP,Sheng06}), becoming   a paradigm of modern physics.

Anderson localization results from the interference of multiply scattered waves, manifesting itself in a most pronounced way in one-dimensional systems~\cite{LGP,FG}, in which all states become localized~\cite{Mott} so that the envelope of each state decays exponentially away from a randomly located localization center \cite{LGP}. The rate of this decay is non-random and is
called the Lyapunov exponent, $\gamma $, the reciprocal of which determines the size of the area of localization.

In a finite, but sufficiently long, disordered sample, the localization manifests itself in the fact that the frequency dependent transmission amplitude is (typically) an exponentially decreasing function of the sample size. The average of this decrement is a size-dependent quantity, whose inverse ({\em i.e.}, reciprocal)  is termed the \textit{transmission length,} $l_{N}$. In the limit as the sample becomes of infinite
length, the decrement tends to a constant non-random value. The reciprocal
of this value determines another characteristic spatial scale of the
localized regime, which is the  \textit{localization length,} $l$. It is
commonly accepted in both the solid state physics and optical communities, that the inverse of the Lyapunov exponent, $\gamma ^{-1},$ and the localization
length, $l$ are always equal. While this is true for media with a continuous
spatial distribution of the random dielectric constant, in the case of randomly layered samples, the situation, as we show in this paper, is more complicated. In particular, the inverse of the Lyapunov exponent
by itself, calculated, for example, in Ref. \cite{Israilev}, does not provide
comprehensive information about the transport properties of disordered media.  Furthermore, it is unlikely that it can be measured directly, at least in the optical regime.

The first study of localization in metamaterials was presented in Ref.~\cite
{Lyapunov} where  wave transmission through an alternating sequence of
air layers and metamaterial layers of random thicknesses was studied.
Localized modes within the gap were observed and delocalized modes were
revealed despite the one-dimensional nature of the model. A more general
model of alternating sequences of right- and left-handed layers with random
parameters was studied in Ref.~\cite{we}. There, it has been shown that in mixed stacks (M-stacks) with fluctuating refractive indices, localization of
low-frequency radiation was dramatically suppressed so that the localization
length exceeded that for homogeneous stacks (H-stacks), composed solely of
right- or left-handed slabs, by many orders of magnitude and scaling as $l\propto \lambda ^{6}$ or even higher powers of wavelength (in what follows
we refer to this result as the $\lambda ^{6}$ anomaly), in contrast to the
well-known dependence $l\propto \lambda ^{2}$ observed in H-stacks~\cite%
{Baluni}. As noted in Ref. \cite{we}, a possible physical explanation of
this is the suppression of phase accumulation in M-stacks, related to the
opposite signs of the phase and group velocities in left- and right-handed
layers. Scaling laws of the transmission through a similar mixed
multilayered structure were studied in Ref.~\cite{random}. There, it was shown
that the spectrally averaged transmission in a frequency range around the
fully transparent resonant mode decayed with the number of layers much more
rapidly than in a homogeneous random slab. Localization in a disordered
multilayered structure comprising alternating random layers of two different
left-handed materials was considered in Ref.~\cite{single}, where it was
shown that within the propagation gap, the localization length was shorter
than the decay length in the underlying periodic structure, and the opposite
of that  observed in the corresponding random structure of right-handed
layers.

In this paper, we study the wave transmission through disordered M- and
H-stacks of a finite size composed of a weakly scattering right- and
left-handed layers. In the framework of the weak scattering approximation
(WSA), we have developed a unified theoretical description of the
transmission and localization lengths over a wide wavelength
range, allowing us to explain the pronounced difference in the
transmission properties of M- and H-stacks at long wavelengths.

When both refractive index and layer thickness of the mixed stack are
random, the transmission length in the long wavelength part of the localized regime exhibits a quadratic power law dependence on  wavelength with different constants of proportionality for mixed and homogeneous stacks. Moreover, in the localized regime, the transmission length of a mixed stack differs from the reciprocal of the Lyapunov exponent of the corresponding infinite stack  (to the best of our knowledge, in all one-dimensional disordered systems studied till now these two quantities always coincide).

%---a result which, to the best of our knowledge, has not previously been found in any one dimensional %disordered system.

Both M- and H-stacks demonstrate a rather narrow crossover from the localized to the ballistic regime. The H-stack in the near ballistic region, and the M-stack in the ballistic region are weakly scattering disordered stacks, while in the far ballistic region, the H-stack transmits radiation as an effectively uniform medium.

We also consider the effects of loss and show that absorption dominates the
effects of disorder at very short and very long wavelengths. The crossover
region is particularly sensitive to losses, so that even small absorption
suppresses oscillations in the transmission length as a function of
frequency.

All of the  theoretical results mentioned above are
confirmed by, and are shown to be in excellent agreement with,
the results of extensive numerical simulation.

In M-stacks with only refractive index disorder, Anderson localization and
transmission resonances are effectively suppressed and the crossover region
between the localized and ballistic regimes is orders of magnitude greater. A
more detailed study of the $\lambda ^{6}$-anomaly shows that the genuine
wavelength dependence of the transmission length is not described by any
power law and rather is non-analytic in nature.

In what follows, Sec. \ref{sec:mod} presents a detailed description of our model. Section \ref{sec:anal} is devoted to the analytical studies of the problem, while the results of numerical simulations and a discussion of these are presented in Sec.~\ref{sec:num}.

%%%%%%%%%%%%%%%%%%%%%%%%%%%%%%%%%%%%%%%%%%%%%%%%%%%%%%

\section{Model}

\label{sec:mod} %%%%%%%%%%%%%%%%%%%%%%%%%%%%%%%%%%%%%%%%%%%%%%%%%%%%%%%%

\subsection{Mixed and homogeneous stacks}

\label{subsec:mixandhomo}
%%%%%%%%%%%%%%%%%%%%%%%%%%%%%%%%%%%%%%%%%%%%%%%%%%%%%%%%%%%

We consider a one-dimensional alternating M-stack, as shown in Fig.~\ref%
{Fig1}. It comprises disordered mixed left- ($L$) and right- ($R$) handed
layers, which alternate over its length of $N$ layers, where $N$ is an even
number. The thicknesses of each layer are independent random values with the
same mean value $d$. In what follows, all quantities with the dimension of
length are measured in units of $d$. In these units, for the thicknesses of
a layer we can write

\begin{equation}
d_{j}=1+\delta _{j}^{(d)},  \notag  \label{d}
\end{equation}%
where the fluctuations of the thickness, $\delta _{j}^{(d)},$ $j=1,2,...$
are zero-mean independent random numbers.
%%%%%%%%%%%%%%%%%%%%%%%%%%%%%%%%%%%%%%%%%%%%%%%%%%%%%%%%%%%%%%%%%%%%%%%%%%%%%%%%%%%%%%%%
\begin{figure}[h]
\scalebox{0.30}{\includegraphics{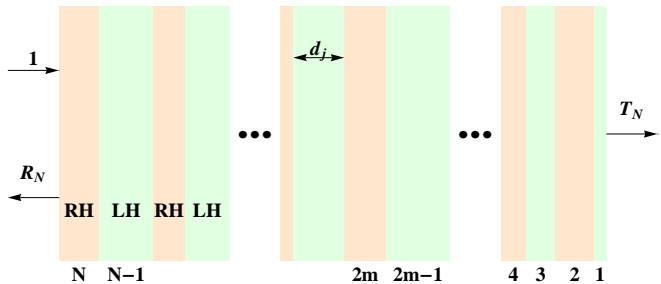}}
\caption{(Color online) Structure geometry.}
\label{Fig1}
\end{figure}
%%%%%%%%%%%%%%%%%%%%%%%%%%%%%%%%%%%%%%%%%%%%%%%%%%%%%%%%%%%%%%%%%%%%%%%%%%%%%%%%%%%%%%%%
We take the magnetic permeability for right-handed media to be $\mu_j=1$ and
for metamaterials to be $\mu_j = -1$, while the dielectric permittivity is

\begin{equation}
\varepsilon _{j}=\pm (1+\delta _{j}^{(\nu )}\pm i\sigma _{j})^{2},  \notag
\label{eps}
\end{equation}%
where the upper and lower signs respectively correspond to normal
(right-handed) and metamaterial (left-handed) layers. The refractive index
of each layer is then

%[nu-1]
\begin{equation}
\nu _{j}=\pm (1+\delta _{j}^{(\nu )})+i\sigma _{j},  \notag  \label{nu-1}
\end{equation}%
where all $\delta _{j}^{(\nu )}$ and absorption coefficients of the slabs, $%
\sigma _{j}\geq 0$, are independent random variables. With this, the
impedance of each layer relative to the background (free space) is

\begin{equation}  \label{imp}
Z_j=\sqrt{\mu_j/\varepsilon_j}=1/(1+\delta^{(\nu)}_j\pm i\sigma_j),  \notag
\end{equation}
with the same choice of the sign.

We begin with the general case when both types of disorder (in refractive
index and in thickness) are present. Two particular cases, each with only
one type of disorder, are rather different. In the absence of absorption,
the M-stack with only thickness disorder is completely transparent, a
consequence of $Z_{j}\equiv 1$. However, the case of only refractive-index
disorder is intriguing because, as  is shown below, such mixed stacks
manifest a dramatic suppression of Anderson localization in the long wave
region \cite{we}.

Although localization in disordered H-stacks with right-handed layers has
been studied by many authors  \cite{Baluni,deSterke,Asatrian2,Sheng06,Luna}, here we consider this problem in its most general form and show that
the transmission properties of disordered H-stacks are qualitatively the
same for stacks comprised of either solely left- or right-handed layers.
%%%%%%%%%%%%%%%%%%%%%%%%%%%%%%%%%%%%%%%%%%%%%%%%%%%%%%%%%%%

\subsection{Transmission: localized and ballistic regimes}

\label{subsec:translocbal}
%%%%%%%%%%%%%%%%%%%%%%%%%%%%%%%%%%%%%%%%%%%%%%%%%%%%%%%%%%%
We introduce the transmission length $l_{N}$ of a finite random
configuration as

%[LL-0]
\begin{equation}
\frac{1}{l_{N}}=-\left\langle \frac{\ln |T_{N}|}{N}\text{ }\right\rangle ,
\label{LL-0}
\end{equation}%
where $T_{N}$ is the random transmission coefficient of a sample of the length $N.$ As a consequence of the self-averaging of $\ln |T_{N}|/N$,

%[LL-3]
\begin{equation}
\lim_{N\rightarrow \infty }\frac{\ln |T_{N}|}{N}=\lim_{N\rightarrow \infty }%
\frac{1}{l_{N}}=\frac{1}{l}.  \label{LL-3}
\end{equation}%
This means that for a sufficiently long stack (in the localized regime) the
transmission coefficient is exponentially small $\left\vert T_{N}\right\vert
\sim \exp {(-\emph{N/l})}$.

In what follows, we consider stacks composed of layers with low dielectric
contrasts, {\it i.e}, $|\delta ^{\nu ,d}|\ll 1,$ so that the Fresnel reflection
coefficients of each interface, and of each layer, are much smaller than $1.$
Here,  a thin stack comprising a small number of layers is almost
transparent. In this, the ballistic regime,  the transmission
length takes the form

%[LL-21]
\begin{equation}
\frac{1}{l_{N}}\approx \frac{\langle |R_{N}|^{2}\rangle }{2N},  \label{LL-21}
\end{equation}%
involving the average reflection coefficient  \cite{Rytov}, which is valid in the case of lossless structures.  This follows directly from Eq. (\ref{LL-0}) by virtue of the conservation relationship, $|R_N|^2 + |T_N|^2 = 1$.  Thus, in the ballistic regime,

%[LL-31]
\begin{equation}
\langle |R_{N}|^{2}\rangle \approx \frac{2N}{b},\ \ \ N\ll b.  \label{LL-31}
\end{equation}%
where the length $b$ in this equation is termed the \textit{ballistic length}.

Accordingly, in studies of the transport of the classical waves in one-dimensional
random systems, the following spatial scales arise in a natural way:
\begin{itemize}
\item $l_{N}$ --- the transmission length of a finite sample (\ref{LL-0}),

\item $l$ --- the localization length (\ref{LL-3}), and

\item $b$ --- ballistic length (\ref{LL-31}).
\end{itemize}

\noindent Note that in the case of absorbing stacks ($\sigma _{j}=\sigma >0$%
), the right hand side of Eq. (\ref{LL-0}) defines the attenuation length, $l_{\mathrm{att}}$,  which incorporates the effects of both
disorder and absorption.

In what follows, we show that contrary to commonly accepted belief, the
quantities $\gamma ^{-1}$, $l$, and $b$ are not necessarily equal, and,
under certain situations, can  differ noticeably from each other.

In this paper, we  study mainly the transmission length defined above by Eq. %
\ref{LL-0}. This quantity is very sensitive to the size of the system and
therefore is best suited to the description of the transmission properties in
both the localized and ballistic regimes. More precisely, the transmission
length coincides either with the localization length or with the ballistic
length, respectively in the cases of comparatively thick (localized regime)
or comparatively thin (ballistic regime) stacks.  That is,

%[LL-4]
\begin{equation}
l_{N}\approx \left\{
\begin{array}{ccc}
l &  & N\gg l \\
&  &  \\
b &  & N\ll b.%
\end{array}%
\right. .  \notag
\end{equation}%
Another argument supporting our choice of the transmission length as the
subject of investigation is that it can be found directly by standard
transmission experiments, while measurements of the Lyapunov exponent call
for a much more sophisticated arrangement.

\section{Analytical studies}

\label{sec:anal} %%%%%%%%%%%%%%%%%%%%%%%%%%%%%%%%%%%%%%

\subsection{Weak scattering approximation}

\label{subsec:weak} %%%%%%%%%%%%%%%%%%%%%%%%%%%%%%%%%%%%%%%%

The theoretical analysis involves the calculation of the transmission
coefficient using a recursive procedure. Consider a stack which is a sequence
of $N$ layers enumerated by index $n$ from $n=1$ at the rear of the stack
through to $n=N$ at the front. The total transmission  ($T_{n})$ and
reflection  ($R_{n}$) amplitude coefficients of the stack satisfy the recurrence relations
%[rec1],[rec2]
\begin{eqnarray}
T_{n} &=&\frac{T_{n-1}t_{n}}{1-R_{n-1}r_{n}},  \label{rec1} \\
R_{n} &=&r_{n}+\frac{R_{n-1}t_{n}^{2}}{1-R_{n-1}r_{n}}  \label{rec2}
\end{eqnarray}
for $n=2,\ldots ,N$, in which both the input and output media are free
space. In Eqs (\ref{rec1}) and (\ref{rec2}), the amplitude transmission ($t_{j}$) and the reflection ($r_{j}$) coefficients  of a single layer are given by

%[r],[t]
\begin{eqnarray}  \label{t}
r_{j} &=&\frac{\rho _{j}(1-e^{2i\beta _{j}})}{1-\rho _{j}^{2}e^{2i\beta _{j}}%
}, \\
t_{j} &=&\frac{(1-\rho _{j}^{2})e^{i\beta _{j}}}{1-\rho
_{j}^{2}e^{2i\beta_{j}}}.
\end{eqnarray}%
Here, $\beta _{j}=kd_{j}\nu _{j}$, $k=2\pi /\lambda ,$ and $\lambda$ denotes
the dimensionless free space wavelength. While the sign of the phase shift
across each slab $\mathrm{Re}\,(\beta _{j})$ varies according to the
handedness of the material, the Fresnel interface coefficient $\rho _{j}$
given by

%[ro]
\begin{equation}  \label{ro}
\rho _{j}=\frac{Z_{j}-1}{Z_{j}+1},
\end{equation}%
depends only on the relative impedance of the layer $Z_{j}$, a quantity
whose real part is positive, irrespective of the handedness of the material.

Equations (\ref{rec1})-(\ref{ro}) are general and provide an \textit{exact}
description of the system and will be used later for direct numerical
simulations of its transmission properties.

It was mentioned previously that we consider the special case of weak scattering for which the reflection from a single layer is small. {\em i.e.},  $\left\vert r_{j}\right\vert \ll 1$. This occurs either for weak disorder, or for strong disorder provided that the wavelength is sufficiently long. The transmission length then follows from

%[RealPart]
\begin{equation}  \label{RealPart}
\ln |T_{N}|^{2}=2\mathrm{Re}\ln T_{N},
\end{equation}
and requires the following first order approximations derived from
Eqs.~(\ref{rec1}) and (\ref{rec2}):

%[rec3],[rec4]
\begin{eqnarray}
\ln T_{n} &=&\ln T_{1,n-1}+\ln t_{n}+R_{n-1}r_{n},  \label{rec3} \\
R_{n} &=&r_{n}+R_{n-1}t_{n}^{2}.\label{rec4}
\end{eqnarray}%
In deriving Eq. (\ref{rec4}), we omit the first-order term $%
R_{n-1}^{2}t_{n}^{2}r_{n}$ since it contributes only to the second order of $%
\ln T_{n}$ already after the first iteration. Then, by summing up
logarithmic terms (\ref{rec3}), we obtain

%[lnT1]
\begin{equation}  \label{lnT1}
\ln T_{N}=\sum_{j=1}^{N}\ln t_{j}+
\sum_{m=2}^{N}\sum_{j=m}^{N}r_{j-m+1}r_{j}\prod_{p=j-m+2}^{j-1}t_{p}^{2}.
\end{equation}

This equation enables us to derive a general expression for the transmission
length $l_{T}(N)$ which is valid in all regimes (see the next Section).
However, the ballistic length $b$, according to Eq. (\ref{LL-31}) (and the average reflection coefficient as well), is determined only by the total reflection amplitude $R_{N}$ in the case of lossless strutures. In the ballistic regime,
this amplitude coefficient, to the necessary accuracy, is given by

%[Refl]
\begin{equation}  \label{Refl}
R_{N}=\sum_{j=1}^{N}r_{j}.
\end{equation}
%%%%%%%%%%%%%%%%%%%%%%%%%%%%%%%%%%%%%%

\subsection{Mixed stack}

\label{subsec:mixstack} %%%%%%%%%%%%%%%%%%%%%%%%%%%%%%%%%%%%%%%%

\subsubsection{General Approach}

\label{subsubsec:gen} %%%%%%%%%%%%%%%%%%%%%%%%%%%%%%%%%%%%%%%%%%%%%%%%%%
From here on, we assume that the random variables $\delta _{j}^{(\nu )}$ of left-handed or right-handed layers, $\delta _{j}^{(d)}$, and $\sigma _{j}$
are identically distributed according to the corresponding probability
density functions. This enables us to express all of the required quantities
via the transmission and reflection amplitudes of a single right-handed or left-handed layer, $t_{r,l}$, $r_{r,l}$, and also to calculate easily all of the necessary ensemble averages.

The average of the first term in Eq. (\ref{lnT1}) can be written as

%[lnTA]
\begin{equation}  \label{lnTA}
\left< \sum_{j=1}^{N}\ln t_{j}\right> =\frac{N}{2}\langle \ln t_{r}\rangle +
\frac{N}{2}\langle \ln t_{l}\rangle .  \notag
\end{equation}
Next, we split the second term of Eq. (\ref{lnT1}) into two parts

%[lntT10]
\begin{equation}  \label{lnT10}
\sum_{m=2}^{N}\sum_{j=m}^{N}r_{j-m+1}r_{j}\prod_{p=j-m+2}^{j-1}t_{p}^{2}=N%
\mathcal{R}_{1}+N\mathcal{R}_{2},  \notag
\end{equation}%
where

%[Rterm1],[Rterm2]
\begin{eqnarray}
\mathcal{R}_{1} &=&\frac{1}{N}\sum_{m=1}^{N/2}\sum_{j=2m}^{N}r_{j-2m+1}r_{j}%
\prod_{p=j-2m+2}^{j-1}t_{p}^{2},  \notag  \label{Rterm2} \\
\mathcal{R}_{2} &=&\frac{1}{N}\sum_{m=1}^{N/2}\sum_{j=2m-1}^{N}r_{j-2m}r_{j}%
\prod_{p=j-2m+1}^{j-1}t_{p}^{2},  \notag
\end{eqnarray}%
comprising contributions to the depletion of the transmitted field due to two pass reflections respectively between slabs of different materials ({\em i.e.}, of opposite handedness), and between slabs of the same material ({\em i.e.}, of like handedness).  Averaging these expressions, we obtain

%[Rterm3],[Rterm4]
\begin{eqnarray}  \label{Rterm3}
\langle \mathcal{R}_{1}\rangle & = & A_r A_l \left [ \frac{1}{1-B^2}+\frac{%
(B^{N}-1)(1+B^2)}{N(1-B^2)^2} \right ], \\
\langle \mathcal{R}_{2}\rangle &=& \frac{A_{r}^{2}B_{l}+A_{r}^{2}B_{l}}{2N}
\left [\frac{N}{1-B^2}+\frac{2(B^{N}-1)}{(1-B^2)^2}\right ],  \label{Rterm4}
\end{eqnarray}%
where

%[ref-trans]
\begin{equation}  \label{ref-trans}
A_{\tau }=\langle r_{\tau }\rangle ,\ \ B_{\tau }=\langle
t_{\tau}^{2}\rangle, \ \ \tau =l,r,\ \ B^{2}=B_{l}B_{r}.
\end{equation}
The resulting transmission length is determined by the equation

%[translength-1]
\begin{equation}
-\frac{1}{l_{N}}=\frac{\langle \ln |t_{r}|\rangle +\langle \ln
|t_{l}|\rangle }{2}+\mathrm{Re}\left( \langle \mathcal{R}_{1}\rangle
+\langle \mathcal{R}_{2}\rangle \right) .  \label{translength-1}
\end{equation}%
In the lossless case ($\sigma =0$), the first term on the right hand side of
Eq. (\ref{translength-1}) corresponds to the so-called single-scattering
approximation, which implies that multi-pass reflections are neglected so
that the total transmission coefficient is approximated by the product of
the single layer transmission coefficients, {\em i.e.},

%[class-picture]
\begin{equation}
|T_{N}|^{2}\ \rightarrow \ \prod_{j=1}^{N}|t_{j}|^{2}.  \notag
\label{class-picture}
\end{equation}%
In the case of very long stacks ({\em i.e.}, as the length $N\rightarrow \infty $%
), we can replace the arithmetic mean, $N^{-1}\sum_{j=1}^{N}\ln |t_{j}|$, by
its ensemble average $\langle \ln |t|\rangle .$ On the other hand, in
this limit the reciprocal of the transmission length coincides with the
localization length. Using the energy conservation law, $%
|r_{j}|^{2}+|t_{j}|^{2}=1$, which applies in the absence of absorption, the
inverse single-scattering localization length may be written as

%[class-picture-2]
\begin{equation}
\left( \frac{1}{l}\right) _{ss}=\frac{1}{2}\langle |r|^{2}\rangle
\label{class-picture-2}
\end{equation}%
and is proportional to the mean reflection coefficient of a single random
layer \cite{LGP,Liansky-1995}. The corresponding modification of Eq. (\ref{translength-1}) then reads

%[translength-2]
\begin{equation}
\frac{1}{l_{N}}=\frac{\langle |r_{r}|^{2}\rangle +\langle |r_{l}|^{2}\rangle
}{4}-\mathrm{Re}\left( \langle \mathcal{R}_{1}\rangle +\langle \mathcal{R}%
_{2}\rangle \right) .  \label{translength-2}
\end{equation}%
Here, the first term corresponds to the single-scattering approximation,
while the next two terms take into account the interference of multiply
scattered waves as well as the dependence of the transmission length on the
stack size. Note that Eq. (\ref{translength-2}) is appropriate only for
lossless structures. In the presence of absorption, Eq. (\ref{translength-1}) should be used instead.

%%%%%%%%%%%%%%%%%%%%%%%%%%%%%%%%%%%%%%

\subsubsection{Transmission length}

\label{subsubsec:transm} %%%%%%%%%%%%%%%%%%%%%%%%%%%%%%%%%%%%%%%%
From this point on, we assume that the statistical properties of the
right-handed and left-handed layers are identical. As a consequence of this
symmetry, the following relations hold for any real-valued function $g$ in
either the lossless or absorbing cases:

%[symmetry]
\begin{equation}
\langle g(t_{r})\rangle =\langle g(t_{l})\rangle ^{\ast },\ \ \ \langle
g(r_{r})\rangle =\langle g(r_{l})\rangle ^{\ast }.  \label{symmetry}
\end{equation}%
Therefore, $A_{r}$ and $B_{r}$ are the complex conjugates of $A_{l}$ and $%
B_{l}$, and $B^{2}$ is real quantity, as are both averages $\langle \mathcal{%
R}_{1}\rangle $ and $\langle \mathcal{R}_{2}\rangle $.

Accordingly, as a consequence of the left-right symmetry (\ref{symmetry}), the
transmission length of a M-stack depends only on the properties of a single
right-handed layer and may be expressed in terms of three averaged
characteristics: $\langle r\rangle $, $\langle \ln |t|\rangle $, and $%
\langle t^{2}\rangle $ (in which we omit the subscript $r$).

With these observations, the transmission length of a finite length M-stack
may be cast in the form:

%[FinalG3]
\begin{equation}  \label{FinalG3}
\frac{1}{l_N}=\frac{1}{l}+\left(\frac{1}{b}-\frac{1}{l}\right) f(N,\bar{l}).
\end{equation}
where

%[loc]
\begin{equation}
\frac{1}{l}=-\langle \ln |t|\rangle -\frac{|\langle r\rangle |^{2}+\mathrm{Re%
}\left( \langle r\rangle ^{2}\langle t^{2}\rangle ^{\ast }\right) }{%
1-|\langle t^{2}\rangle |^{2}},  \label{loc}
\end{equation}%
and

%[bal]
\begin{eqnarray}
\frac{1}{b} &=&\frac{1}{l}-\frac{2/\bar{l}}{1-\exp (-2/\bar{l})}\times
\notag  \label{bal} \\
&&\left( \frac{|\langle r\rangle |^{2}+\mathrm{Re}\left( \langle r\rangle
^{2}\langle t^{2}\rangle ^{\ast }\right) }{1-|\langle t^{2}\rangle |^{2}}-%
\frac{|\langle r\rangle |^{2}}{2}\right)
\end{eqnarray}%
are, as we will see below, the inverse localization and inverse ballistic lengths. The
function $f(N,\bar{l})$ is defined as

%[f]
\begin{equation}
f(N,\bar{l})=\frac{\bar{l}}{N}\left[ 1-\exp \left( -\frac{N}{\bar{l}}\right) %
\right] ,  \label{f}
\end{equation}%
and introduces a new characteristic length termed the \emph{crossover length}

%[l]
\begin{equation}  \label{l}
\bar{l}=-\frac{1}{\ln |\langle t^{ 2}\rangle |},
\end{equation}%
which arises in the calculations in a natural way and, as will be
demonstrated below, plays an important role in the theory of
the transport and
localization in one-dimensional random systems.  Equations (\ref{FinalG3})-(\ref{bal}) completely describe the
transmission length of a mixed stack
in the weak scattering approximation.

Obviously, the characteristic lengths $l(\lambda )$, $b(\lambda )$, and $\bar{l}(\lambda )$ appearing in Eq. (\ref{FinalG3}) are functions of
wavelength. Using straightforward calculations, it may be shown that the
first two always satisfy the inequality $l(\lambda )>b(\lambda )$, while, as
we will see, the crossover length is the shortest of the three, {\em i.e.}, $%
b(\lambda )>\bar{l}(\lambda )$ in the long wavelength region.

In the case of a fixed wavelength $\lambda $ and a stack so short that $N\ll
\bar{l}(\lambda )$, the expansion of the exponent in Eq. (\ref{f}) yields $f\rightarrow 1$, in which case the transmission length approaches $b(\lambda)$. Correspondingly, for a sufficiently long stack, $N\gg \bar{l}(\lambda )$,  $f\rightarrow 0$ and the transmission length assumes the value of $l(\lambda )$.

In summary,
\begin{equation}
l_{N}(\lambda )\approx \left\{
\begin{array}{ccc}
l(\lambda ), &  & N\gg \bar{l}(\lambda ), \\
&  &  \\
b(\lambda ), &  & N\ll \bar{l}(\lambda ),%
\end{array}%
\right.   \notag  \label{crossover-1}
\end{equation}%
with the transition between the two ranges of $N$ being
determined by the crossover length.

While in the lossless case, the ballistic regime occurs when the stack is
much shorter than the crossover length ($N\ll \bar{l}(\lambda )$), it is
important to note that, in the localization regime, the opposite inequality is
not sufficient and the necessary condition for localization is $N\gg
l(\lambda )$. In what follows, we consider samples of an intermediate
length, {\em i.e.}, $\bar{l}(\lambda )\ll N\ll l(\lambda )$.

For a M-stack of  fixed size $N,$ the parameter governing the
transmission is the wavelength, and the conditions for the localized and
ballistic regimes should be formulated in the wavelength domain. To do this,
we introduce two characteristic wavelengths, $\lambda _{1}(N)$ and $\lambda
_{2}(N),$ defined by the relations

%[wavelengths-1]
\begin{equation}  \label{wavelengths-1}
N=l(\lambda _{1}(N)),\ \ \ N=\bar{l}(\lambda _{2}(N)).
\end{equation}

It can be shown that the long wavelength region, $\lambda \ll \lambda _{1}(N)
$, corresponds to localization where the transmission length coincides with
the localization length, while in the extremely long wavelength region, $%
\lambda \gg \lambda _{2}(N)$, the propagation is ballistic, with the
transmission length given by the ballistic length $b$. That is,

%[crossover-2]
\begin{equation}
l_{N}(\lambda )\approx \left\{
\begin{array}{ccc}
l(\lambda ), &  & \lambda \ll \lambda _{1}(N), \\
&  &  \\
b(\lambda ), &  & \lambda \gg \lambda _{2}(N).%
\end{array}%
\right.  \label{crossover-2}
\end{equation}%
When $\lambda _{1}(N)<\lambda _{2}(N)$, there exists an intermediate range
of  wavelengths, $\lambda _{1}(N)<\lambda <\lambda _{2}(N)$, which will
be discussed below.

To better understand the physical meaning of the expressions (\ref{loc}) and
(\ref{bal}) for the localization and ballistic lengths, we will consider
ensembles of random configurations in which the fluctuations $\delta
_{j}^{(\nu )}$ and $\delta _{j}^{(d)}$ are distributed uniformly over the
intervals $[-Q_{\nu },Q_{\nu }]$ and $[-Q_{d},Q_{d}]$ respectively, with $\sigma _{j}=0$. The average quantities that arise in Eqs. (\ref{loc}) and (%
\ref{bal}) are presented in the Appendix. These formulae allow for calculations
with an accuracy of order $O(Q_{\nu }^{2})$ for  arbitrary $Q_{d}$. For the
sake of simplicity, we assume that the fluctuations of the refractive index
and thickness are of the same order, {\em i.e.}, $Q_{\nu }\sim Q_{d}$ so that the
dimensionless parameter
\begin{equation*}
\zeta =2\frac{Q_{d}^{2}}{Q_{\nu }^{2}}
\end{equation*}%
is of order of unity. We also neglect the contribution of terms of order higher than $Q_{d}^{2}$.

The short wavelength asymptotic of the localization length is then
%[shortWave]
\begin{equation}
l(\lambda )=\frac{12}{Q_{\nu }^{2}}.  \label{shortWave}
\end{equation}%
In the long wavelength limit, we obtain the following asymptotic forms for
the corresponding single layer averages:
%[expansions-r],[expansions-ln-t],[expansions-t]
\begin{eqnarray}
\langle r\rangle  &\approx &\frac{ikQ_{\nu }^{2}}{6}-\frac{k^{2}Q_{\nu }^{2}%
}{2}-\frac{5ik^{3}Q_{\nu }^{2}}{9},  \label{expansions-r}\\
\langle \ln |t|\rangle  &\approx &-\frac{k^{2}Q_{\nu }^{2}}{6}, \label{expansions-ln-t}\\
\langle t^{2}\rangle  &\approx &1+2ik+\frac{ikQ_{\nu }^{2}}{3}- \notag\\
&&-2k^{2}-\frac{5k^{2}Q_{\nu }^{2}}{3}+\frac{2k^{2}Q_{d}^{2}}{3}\label{expansions-t}.
\end{eqnarray}%
Substitution of these expansions into Eq. (\ref{bal}) yields the long
wavelength asymptotic of the ballistic length
%[bbal]
\begin{equation}
b(\lambda )\approx \frac{3\lambda ^{2}}{2\pi ^{2}Q_{\nu }^{2}}.  \label{bbal}
\end{equation}%

This asymptotic can be calculated directly from Eq. (\ref{LL-31}) with the average reflection coefficient, determined from Eq. (\ref{Refl}), being

%[Refl-mix]
\begin{equation}  \label{Refl-mix}
\left\langle\left| R_N\right| ^2\right\rangle=N\left( \left\langle\left|
r\right| ^2\right\rangle- \left| \left\langle r\right\rangle \right|^2
\right)+ N^2 \left\langle \mathrm{Re} \ r \right\rangle^2.
\end{equation}

The same substitutions into this equation (\ref{Refl-mix}) give the average total
reflection coefficient

%[totalRefl]
\begin{equation}
\left\langle \left\vert R_{N}\right\vert ^{2}\right\rangle \approx \frac{%
Nk^{2}Q_{\nu }^{2}}{3} + \frac{N^2 k^4 Q_\nu^4}{4}.  \label{totalRefl}
\end{equation}%
In the ballistic regime, the final term is negligibly small.  This, together with Eq. (\ref{LL-31}), again results in the value of the ballistic length given in Eq. (\ref{bbal}).

Substituting the long wavelength expansions (\ref{expansions-r})--(\ref{expansions-t}) into Eqs. (\ref{loc}) and (\ref{l}), we derive the following asymptotic forms for the localization length

%[lloc]
\begin{equation}  \label{lloc}
l(\lambda)\approx \frac{3\lambda^2}{2\pi^2Q_{\nu}^2} \ \frac{3+\zeta}{1+\zeta%
},
\end{equation}
and the crossover length

%[crossover-3]
\begin{equation}
\bar{l}(\lambda )\approx \frac{3\lambda^2}{2\pi^2Q_{\nu}^2} \frac{1}{%
4(3+\zeta )}.  \label{crossover-3}
\end{equation}

In seeking to compare the result (\ref{lloc}) for the localization length with the corresponding long wave asymptotic of the reciprocal Lyapunov exponent, it is important to note that these two quantities are defined in different ways. The localization length is defined by Eq. (\ref{LL-3}) via the transmittivity on a realization, while the  Lyapunov exponent describes an exponential growth of the envelope of a currentless solution far from a given point in which the solution has a given value \cite{LGP}.  We have calculated the long wave asymptotic of the M-stack Lyapunov exponent using the well known transfer matrix approach, assuming the same statistical properties of the modulus of dielectric constant and the thickness of both left- and right-handed layers,
%[Lyapunov-3]

\begin{eqnarray}
&&\gamma \approx\frac{\pi ^{2}\overline{d^{2}}}{2\lambda ^{2}}\frac{\overline{%
\epsilon ^{2}}-\overline{\epsilon }^{2}}{\overline{\epsilon }},
\label{Lyapunov-3} \\
\epsilon =&&(1+\delta^{(\nu)})^2,  \ \ \ d=1+\delta^{(d)}.\nonumber
\end{eqnarray}%
In the case of rectangular distributions of the fluctuations of the dielectric constants and thicknesses, this result reduces to
%[Lyapunov-4]

\begin{equation}
\gamma \approx\frac{2\pi ^{2}Q_{\nu }^{2}}{3\lambda ^{2}}=\frac{1}{l}\frac{3+\zeta
}{1+\zeta }>\frac{1}{l}  \label{Lyapunov-4}
\end{equation}%
(we neglected the small corrections proportional to $Q_{d}^{2}\ll 1$).
Thus, the disordered M-stack in the long wavelength region presents a unique example of a one-dimensional disordered system in which the localization length differs from the reciprocal of the Lyapunov exponent.

The long wavelength asymptotic of the Lyapunov exponent $\gamma \propto
\lambda ^{-2}$ and the asymptotic of the localization length, $%
l\propto \lambda ^{2}$, Eq. (\ref{lloc}) have rather clear physical
meaning. Indeed, in the limit $\lambda \rightarrow \infty $, the propagating wave is insensitive to disorder since the disorder is effectively averaged over distances of the order of the wavelength. This means that Anderson localization is absent and the Lyapunov exponent $\gamma \propto l^{-1}$\ vanishes. For large but finite wavelengths ({\em i.e.}, small wavenumbers $k=2\pi /\lambda )$, the Lyapunov exponent is small and, assuming that its dependence on the wavenumber is analytic, we can expand it in powers of $k$. This expansion
commences with a term of order $k^{2}$ since the Lyapunov exponent is real
and the wavenumber enters the field equations in the form $(i k)$.
Accordingly, in the long wavelength limit, $\gamma \propto k^{2}$ and so $%
l\propto \lambda ^{2}$.

%This behaviour of the Lyapunov exponent does not depend on the handness of the %layers (compare the latter equation with the corresponding long wave %asymptotic obtained in Refs. [\onlinecite{Ping91,single}] for homogeneous %stacks composed of only positive or only negative layers). However, it %crucially depends on the propagating character of the field within given %wavelength region. When this region is within the gap of the propagation %spectrum of the effective ordered medium, the frequency dependence of the %Lyapunov exponent differs from $\propto\lambda^{-2}$. It takes place e.g., for %a stack composed of single negative layers, where only one of two %characteristics $\varepsilon$ and $\mu$ is negative \cite{single}.

The behavior of the Lyapunov exponent given by (\ref{Lyapunov-3})  does not depend on the handness of the layers;  compare this equation   with the corresponding long wave asymptotic obtained in Refs. \cite{Ping91,single} for homogeneous stacks composed of only positive or only negative layers. However, it crucially depends on the propagating character of the field within a given wavelength region. When this region is within the gap of the propagation spectrum of the effective ordered medium, the frequency dependence of the Lyapunov exponent differs from  $ \sim\lambda^{-2}$. It takes place, for example, for a stack composed of single negative layers, where only one of two characteristics  ($\varepsilon$, $\mu$) is negative \cite{single}.

The main contribution to the ballistic length (\ref{bbal}) is due to the
final term in the right hand side of Eq. (\ref{bal}), which corresponds to
the single-scattering approximation discussed at the end of the Sec.~\ref%
{subsubsec:gen}. While the ballistic length follows from the single
scattering approximation, the calculation of the localization length (\ref%
{lloc}) is more complex and requires that  interference due to the
multiple scattering of waves must be taken into account.

In the case under consideration ($Q_{\nu }\sim Q_{d}$ and $\zeta \sim 1$), the
two characteristic wavelengths $\lambda _{1}(N)$ and $\lambda _{2}(N)$ (\ref{wavelengths-1}) for an M-stack of a fixed length size $N$ take the form

%[wavelengths-3]
\begin{eqnarray}
\lambda _{1}(N) &=&\pi Q_{\nu }\sqrt{\frac{2N}{3}\frac{1+\zeta }{3+\zeta }},
\label{wavelengths-3} \\
\lambda _{2}(N) &=&2\pi Q_{\nu }\sqrt{2N\left( 1+\frac{\zeta }{3}\right) },
\end{eqnarray}%
and are of the same order of magnitude. This means that the transmission
length $l_{N}$ coincides with the localization length (\ref{lloc}) for $N\gg
l(\lambda )$), and with the ballistic length (\ref{bbal}) in the case when $%
N\ll \bar{l}(\lambda )$.

The crossover between the localized and ballistic regimes occurs for $%
\lambda _{1}(N)\lesssim \lambda \lesssim \lambda _{2}(N)$, with the two
bounds being proportional to $Q_{\nu }\sqrt{N}$ and thus growing as $\sqrt{N}
$. For long stacks, the transmission length in the crossover region can be
described with high accuracy by the general equations (\ref{FinalG3}) and (\ref{f}), where the ballistic length $b(\lambda )$, the localization length $l(\lambda)$, and the crossover length $\bar{l}(\lambda )$ are replaced by their asymptotic forms (\ref{bbal})--(\ref{crossover-3}). In the case of solely
refractive index disorder ($Q_{d}\rightarrow 0)$ when all thicknesses are
set to unity, the ballistic length and the short wavelength asymptotic of
the localization length coincide with the limiting values of their
counterparts for an M-stack with both refractive index and thickness
disorder (\ref{bbal}) and (\ref{shortWave}). However, the corresponding
limiting values of the long wavelength localization length and crossover
lengths have nothing to do with the genuine behaviour of the transmission
length (see Sec.~\ref{subsubsec:suppress}). This means that the weak
scattering approximation fails to describe the long wavelength asymptotics
of the transmission length in both the localization and crossover regions.
This is discussed in greater detail below.

Eqs. (\ref{Rterm3})--(\ref{translength-1}) (or Eqs. (\ref{FinalG3})--(\ref%
{l})) in the symmetric case) completely determine the behaviour of the
transmission length for a mixed stack composed of weakly scattering layers.
Although $\bar{l}$ has been introduced as a crossover length, the entire
region $N\gg \bar{l}$ does not necessarily support localization.
Correspondingly, the ballistic regime may exist outside the region $N\ll
\bar{l}.$\

All of the results obtained under the symmetry assumption (\ref{symmetry}) are
qualitatively valid in the general case. Indeed, the existence of the
crossover (\ref{crossover-2}) is related to the exponential dependence in
Eq. (\ref{f}) with a real and positive crossover length $\bar{l}$. When the
assumption of symmetry no longer holds, this length takes a complex value.
However, the quantity $B$ in Eq. (\ref{Rterm4}) satisfies (by its definition in Eq. (\ref{ref-trans})) the evident inequality $|B|<1$. Therefore, the corresponding analogue of the function $f$ (\ref{f}) preserves all necessary limiting properties (see the next Sec.~\ref{subsec:hom}).

\subsubsection{Homogeneous stacks}

\label{subsec:hom} %%%%%%%%%%%%%%%%%%%%%%%%%%%%%%%%%%%%%%%%%%%%%%%%%%%%
In this Section, we consider a H-stack composed entirely of normal material
layers noting that the behaviour of a H-stack of metamaterial (left-handed)
layers alone is exactly the same, a result which may be obtained directly
from Eq. (\ref{translength-1}) by replacing each $l$ by $r$, after which any
reference to the index $r$ may be omitted. The transmission length of an
H-stack is then

%[transhom]
\begin{equation}  \label{transhom}
\frac{1}{l_N} = \frac{1}{l}+\frac{1}{N} \ \mathrm{Re }\left[\langle r
\rangle^2 \frac{1-\langle t^2\rangle^N}{(1-\langle t^2\rangle)^2}\right],
\end{equation}
where the inverse localization length $l$ is

%[lochom]
\begin{equation}  \label{lochom}
\frac{1}{l}= -\langle \ln|t| \rangle - \mathrm{Re }\frac{\langle r \rangle^2%
}{1-\langle t^2\rangle}.
\end{equation}

Now we consider a H-stack composed of  weakly scattering layers. To
simplify the discussion, we consider only refractive index disorder ({\em i.e.}, $%
Q_{d}=0$). In this case, the asymptotic behavior of the localization length
in the short and long wavelength limits are

%[asympt-H]
\begin{equation}
l(\lambda )=\left\{
\begin{array}{ccc}
\displaystyle{\frac{12}{Q_{\nu }^{2}}}, &  & \lambda \rightarrow 0, \\
&  &  \\
\displaystyle{\frac{3\lambda ^{2}}{2\pi ^{2}Q_{\nu }^{2}}}, &  & \lambda
\rightarrow \infty.
\end{array}
\right.
\label{asympt-H}
\end{equation}

The main contribution to the localization length is related to the first
term in Eq. (\ref{lochom}). Thus, the localization length of the H-stack in
the long wavelength region is described  completely by the single scattering
approximation and coincides with the ballistic length (\ref{bbal}) of the
M-stack.

Using the transfer matrix approach, we can also calculate  the long wavelength
asymptotic of the Lyapunov exponent  for a  H-stack . It is described by the same
equation (\ref{Lyapunov-3}) as the asymptotic for the M-stack, thus coinciding
with the asymptotic of the reciprocal Lyapunov exponent. In  recent work
\cite{Israilev}, this coincidence was established analytically in a wider
spectral region. However, the numerical calculations intended to confirm
this result are rather unconvincing. Indeed, the numerically
obtained plots demonstrate strong fluctuations (of the same order as the mean value)  of the calculated quantity, while the genuine Lyapunov exponent
is non-random and should be smooth without any additional ensemble averaging
mentioned by the authors.

If we cast $\langle t^{2}\rangle ^{N}$ in the second term of Eq. (\ref{transhom}) in the form $\exp \left( \displaystyle{N\ln \langle t^{2}\rangle}\right) $ we see that the crossover length of the H-stack is

%[crossover-H-1]
\begin{equation}  \label{crossover-H-1}
\bar{l}= \left|\ln\langle t^2\rangle\right|^{-1},  \notag
\end{equation}
with its long wavelength asymptotic, according to Eq. (\ref{expansions-t}),
being

%[crossover-H-2]
\begin{equation}
\bar{l}(\lambda )=\frac{\lambda }{4\pi }.  \label{crossover-H-2}
\end{equation}%
Here, the crossover length is proportional to the wavelength, in stark
contradistinction to the situation for M-stacks, in which the crossover
length is proportional to $\lambda ^{2}$.

To consider this further, we define the characteristic wavelengths $%
\lambda_1(N)$ and $\lambda_2(N)$ by the expressions (\ref{wavelengths-1}).
For a H-stack, these lengths are

%[crossover-H-3]
\begin{eqnarray}  \label{crossover-H-3}
\lambda_1(N)&=&\pi Q_{\nu}\sqrt{\frac{2N}{3}},  \notag \\
\lambda_2(N)&=&4\pi N.
\end{eqnarray}
Evidently, the second characteristic wavelength is always much larger than
the first  $\lambda_2(N) \gg \lambda_1(N)$. As a consequence, the long
wavelength region, where the ballistic regime is realized, can be divided
into two subregions. The near subregion (moderately long wavelengths) is
bounded by the characteristic wavelengths

%[crossover-H-4]
\begin{equation*}
\lambda _{1}(N)\lesssim \lambda \lesssim \lambda _{2}(N).
\end{equation*}%
The main contribution to the ballistic length in the near ballistic region, $%
b_{n},$ is due to the first term in Eq. (\ref{lochom}). Thus the
ballistic length $b_{n}(\lambda)$ has the same wavelength dependence as the localization
length $l(\lambda )$ (\ref{asympt-H})

%[bal-H-1]
\begin{equation*}
b_{n}(\lambda )=\displaystyle{\frac{3\lambda ^{2}}{2\pi ^{2}Q_{\nu }^{2}}} ,
\end{equation*}%
and is well described by the single scattering approximation (\ref%
{class-picture-2}). For a H-stack , the transition from the localized to
the ballistic regime at $\lambda \sim \lambda _{1}(N)$ is not accompanied by
any change of the wavelength dependence of the transmission length. This
change can occur at much longer wavelengths $\lambda \sim \lambda _{2}(N)$
in the far long wavelength subregion.

To derive the ballistic length $b_{f}$ in the far long wavelength region, we may proceed in a similar manner to that outlined in the case of a  M-stack and expand the exponent $\langle t^{2}\rangle ^{N}=\exp \left( \displaystyle{N\ln \langle
t^{2}\rangle }\right) $ in Eq. (\ref{transhom}). However, because $\langle t^{2}\rangle$ in this expression is complex, the situation is more complicated than was the case for the M-stack. In particular, the first two terms of the expansion do not
contribute to the ballistic length. Taking account of the second order leads
to the following expression for the ballistic length, $b_{f}(\lambda ),$ in
the far long wavelength region

%[bal-H-2]
\begin{equation}
\frac{1}{b_{f}(\lambda )}=\frac{2\pi ^{2}Q_{\nu }^{2}}{3\lambda ^{2}}+\frac{%
N\pi ^{2}Q_{\nu }^{4}}{18\lambda ^{2}}.  \label{bal-H-2}
\end{equation}%

In the case of a relatively short H-stack $NQ_{\nu }^{2}/12\ll 1,$ the
contribution of the first term in the right hand side of this equation dominates, and hence the
transition from the near to the far subregions is not accompanied by any
change in the analytical dependence on the wavelength.  The ballistic length is thus described by the same wavelength dependence over the entire ballistic
region

%[bal-H-3]
\begin{equation}
b(\lambda )=\frac{3\lambda ^{2}}{2\pi ^{2}Q_{\nu }^{2}},\ \ \ \lambda
_{1}(N)\ll \lambda .  \label{bal-H-3}
\end{equation}%

For sufficiently long H-stacks $NQ_{\nu }^{2}/12\gg 1,$ in the far long wavelength region, the second term is dominant and so

%[bal-H-4]
\begin{equation}
b_{f}(\lambda )=\frac{18\lambda ^{2}}{N\pi ^{2}Q_{\nu }^{4}}.
\label{bal-H-4}
\end{equation}%
Thus, the wavelength dependence of the ballistic length of a sufficiently long H-stack is

%[bal-H-5]
\begin{equation}
b(\lambda )=\left\{
\begin{array}{ccc}
\displaystyle{
\frac{3\lambda ^{2}}{2\pi ^{2}Q_{\nu }^{2}}
}
, &  & \lambda _{1}(N)\lesssim \lambda \lesssim \lambda
_{2}(N), \\
&  &  \\
\displaystyle{
\frac{18\lambda ^{2}}{N\pi ^{2}Q_{\nu }^{4}}
}
, &  & \lambda _{2}(N)\lesssim \lambda .
\end{array}%
\right.   \label{bal-H-5}
\end{equation}%

The same result for the ballistic length also follows from Eq. (\ref{LL-31}) with  Eq. (\ref{Refl}), in the case of a homogeneous stack, yielding

%[Refl-hom]
\begin{equation}
\left\langle \left\vert R_{N}\right\vert ^{2}\right\rangle =N\left(
\left\langle \left\vert r\right\vert ^{2}\right\rangle -\left\vert
\left\langle r\right\rangle \right\vert ^{2}\right) +N^{2}\left\vert
\left\langle r\right\rangle \right\vert ^{2}.  \label{Refl-hom}
\end{equation}%
In the long wave limit this leads to

%[bal-H-6]
\begin{equation}
\left\langle \left\vert R_{N}\right\vert ^{2}\right\rangle =\frac{%
Nk^{2}Q_{\nu }^{2}}{3}+\frac{N^{2}k^{2}Q_{\nu }^{4}}{36} ,\ \ \
\lambda _{1}(N)\ll \lambda,   \label{bal-H-6}
\end{equation}%

The final term in Eq. (\ref{Refl-hom}) differs from the final
term in Eq. (\ref{Refl-mix}) and, therefore, in contrast to the
M-stack, its contribution to the total reflection
coefficient can be of the order of, or larger than, that of
the first term.  This together with Eq. (\ref{LL-31})  is equivalent to  the result in Eqs. (\ref{bal-H-3}) and (\ref{bal-H-5}), and is applicable to short and long stacks respectively.

The far long wavelength ballistic asymptotic (\ref{bal-H-4}) has a simple physical interpretation. Indeed, in this subregion, the wavelength essentially exceeds the stack size and so we may consider the stack as a single weakly
scattering uniform layer with an effective dielectric permittivity  $\varepsilon_{\rm eff}.$ In this case, the ballistic length of the stack according to Eq.(\ref{LL-31}) is
%[effective-1]
\begin{equation}  \label{effective-1}
b_{f}=\frac{2N}{|R_{\rm eff}|^2},
\end{equation}
where

%[effective-2]
\begin{equation}
R_{\rm eff}=\frac{ikN}{2}\left( \varepsilon_{\rm eff}-1\right)   \label{effective-2}
\end{equation}%
is the long wavelength form of the reflection amplitude for a uniform right-handed stack
of the length $N$ and constant dielectric permittivity $\varepsilon _{\rm eff}$ (with $\mu=1$).  In the case of a uniform left-handed stack, the value of the reflection coefficient  should be
%[effective-2l]
\begin{equation}
R_{\rm eff}=-\frac{ikN}{2}\left( \varepsilon_{\rm eff}+1\right),   \label{effective-2l}
\end{equation}%
in which the value of $\varepsilon _{\rm eff}$ is now negative (with $\mu=-1$).

To calculate the effective parameters, we use Eq.(\ref{Refl}). Neglecting
small fluctuations of the layer thickness, the single layer reflection
amplitude in the long wavelength limit reads

%[ReflSingle]
\begin{equation}
r_{j}=\frac{ik}{2}\left( \varepsilon_{j}-1\right) .  \label{ReflSingle}
\end{equation}%
The total reflection amplitude for a stack of  length $N$ is
%[effective-3]
\begin{equation}  \label{effective-3}
R_N=\frac{ik}{2}\sum_{j=1}^N\left(\varepsilon_j-1\right)= \frac{ikN}{2}\left(\frac{%
1}{N}\sum_{j=1}^N\varepsilon_j-1\right) ,
\end{equation}
corresponding to the effective dielectric permittivity determined by the expression
%[effective-4]
\begin{equation}
\varepsilon_{\rm eff}=\frac{1}{N}\sum_{j=1}^{N}\varepsilon_{j}  . \label{effective-4}
\end{equation}%
For sufficiently long stacks, the right hand side of this equation
%Eq. (\ref{effective-4})
can be replaced by the ensemble average $\langle \varepsilon _{j}\rangle $ and hence

%[effective-5]
\begin{equation}  \label{effective-5}
\varepsilon_{\rm eff}=\langle \varepsilon_{j}\rangle \approx 1+\frac{Q_{\nu }^{2}}{3}.
\end{equation}%
This expression is similar to that of the two-dimensional case for disordered  photonic crystals  reported in Ref. \cite{Asatryan99}. The same considerations as above, but for a homogeneous stack composed entirely with disordered metamaterial slabs, also lead to an effective value of the permittivity,

\begin{equation}  \label{effective-5b}
\varepsilon_{\rm eff}=\langle \varepsilon_{j}\rangle \approx -\left (1+\frac{Q_{\nu }^{2}}{3}\right).
\end{equation}%

Eq. (\ref{effective-4}), together with Eqs (\ref{effective-1}), (\ref{effective-2}),
%and (\ref%{effective-4}),
leads immediately to the far long wavelength ballistic length (\ref{bal-H-4}).
We emphasize that because of the effective uniformity of the H-stack in the far ballistic region, the transmission length on a single realization is a less fluctuating quantity.  In contrast, transmission length for H-stacks fluctuates strongly  in the near ballistic region, as indeed it does over the entire  ballistic region for M-stacks.

To characterize the entire crossover region between the two ballistic regimes (\ref{bal-H-5}) in greater detail, we return to the general formula (\ref{transhom}), in which we represent ${\langle t^{2}\rangle }^{N}$ as $\exp( N\ln {\langle t^{2}\rangle })$. Then, in accordance with Eq. (\ref{expansions-t}), we can write
\begin{equation}
{\langle t^{2}\rangle }^{N}\approx \exp \{N(2ik-k^{2}Q_{\nu }^{2})\approx
(1-Nk^{2}Q_{\nu }^{2})\exp(2ikN).  \label{brrrr}
\end{equation}
As a consequence,  instead of the result in Eq. (\ref{bal-H-2}), we obtain

%[bal-H-6]
\begin{equation}
\frac{1}{b_{n}(\lambda )}=\frac{2\pi ^{2}Q_{\nu }^{2}}{3\lambda ^{2}}+\frac{%
Q_{\nu }^{4}}{72N}\sin ^{2}{\frac{2\pi N}{\lambda }}.  \label{bal-H-6a}
\end{equation}%
We note that, strictly speaking, the expansion (\ref{brrrr}) is valid inside
the interval $\sqrt{6}\lambda _{1}(N)\ll \lambda \ll \lambda _{2}(N).$

The second term in Eq. (\ref{bal-H-6}) represents standard oscillations of
the reflection coefficient of a uniform  slab of finite size. In the far
long wavelength limit $\lambda \gg \lambda _{2},$ this equation coincides
with (\ref{bal-H-2}). Thus, Eq.~(\ref{bal-H-6}) describes the ballistic
length over practically the whole ballistic region $\lambda \gg \lambda _{1}(N).$
Moreover, taking into account that the long wavelength asymptotic of the localization
length coincides with that of the near long wavelength ballistic length, we see that the right hand side of Eq. (\ref{bal-H-6})   serves as an excellent interpolation formula for
the reciprocal of the transmission length $1/l_{N}(\lambda )$ of a sufficiently long stack ($NQ_{\nu }^{2}/12\gg 1$) over the entire long wavelength region $%
\lambda \gg 1.$ %%%%%%%%%%%%%%%%%%%%%%%%%%%%%%%%%%%%%%%%%%%%%%%%%%%%%%%%%

\subsection{Comparison of the transmission length behaviour in M- and H-
stacks}

\label{subsec:comparison}

Away from the transition regions, the transmission length can exhibit three
types of long wavelength asymptotics described by the right hand sides of
Eqs. (\ref{bbal}), (\ref{lloc}), and (\ref{bal-H-4}). The first (\ref{bbal})
corresponds to the single scattering approximation where the inverse
transmission length is proportional to the average reflection coefficient of
a single random layer. In the absence of absorption, this characterizes both the localization and ballistic
lengths. The second asymptotic form (\ref{lloc}) takes into account the
interference of multiply scattered waves and describes the localization
length. The third asymptotic (\ref{bal-H-4}) corresponds to transmission
through a uniform slab with an effective dielectric constant given by Eq. (%
\ref{effective-5}) and is relevant only in the ballistic regime.

In the case of a M-stack, the first two expressions (\ref{bbal}) and (\ref{lloc})
characterize the ballistic and localized regimes respectively, while the
third asymptotic is never realized in M-stacks. In relatively short H-stacks
(\ref{bal-H-3}), the long wavelength behaviour of the transmission length is
described by the same dependence (\ref{bbal}) in both the localized and
ballistic regimes. Finally, in the case of long H-stacks (\ref{bal-H-4}), the
transmission length follows from Eq. (\ref{bbal}) in the localized and near
ballistic regimes, while in the far ballistic region it is described by the
right hand side of Eq. (\ref{bal-H-4}).

These results predict the existence of a different wavelength dependence of
the transmission length in different wavelength ranges. For M-stacks, the
crossover between them occurs at the wavelength $\lambda _{1}(N)$ (\ref%
{wavelengths-3}) where the size of the stack is comparable with the
localization length $l(\lambda _{1}(N))\approx N$. For long H-stacks, the
crossover occurs when the wavelength becomes comparable to the stack size $%
\lambda \approx N$. Short H-stacks exhibit the same wavelength dependence of
the transmission length in all long wavelength regions, {\em i.e.}, in both
localized and ballistic regimes.

%%%%%%%%%%%%%%%%%%%%%%%%%%%%%%%%%%%%%%%%%%%%%%%%%%%%%%%

\section{Numerical results}

\label{sec:num} %%%%%%%%%%%%%%%%%%%%%%%%%%%%%%%%%%%%%%%%%%%%%%%%%%%%%%%%
The results of the numerical simulations presented below correspond to
uniform distributions of the fluctuations $\delta ^{(d)},$ $\delta ^{(\nu )}$, with widths of $Q_{\nu }$ and $Q_{d}$, respectively, and with a constant
value of the absorption $\sigma $ in each layer.

Results are presented for (a) direct simulations based on the exact
recurrence relations (\ref{rec1}) and (\ref{rec2}); (b) the weak scattering
analysis for the transmission length based on Eqs.~(\ref{FinalG3}) and (\ref%
{transhom}); and (c) asymptotics for short (\ref{shortWave}) and long wavelengths [(\ref%
{bbal}), (\ref{lloc}), and (\ref{bal-H-4})] . For the mean
reflectivity, we used the asymptotic forms (\ref{totalRefl}) and (\ref{bal-H-6}).

In all cases, unless otherwise is mentioned, the ensemble averaging is taken
over $N_{r}=10^{4}$ realizations. The results up to Sec. \ref{subsec:losses}
are for lossless stacks ($\sigma =0$) only.

%%%%%%%%%%%%%%%%%%%%%%%%%%%%%%%%%%%%%%%%%%%%%%%%%%%%%%%%%%%%%%%%%%%%%

\subsection{Refractive-index and thickness disorder}

\label{subsec:refthick}
\begin{figure}[h]
\rotatebox{0}{\scalebox{0.40}{\includegraphics{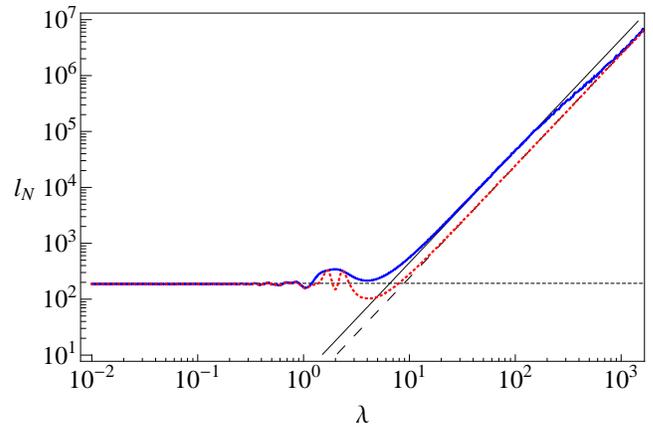}}}
\caption{(Color online) Transmission length $l_{N}$ vs  $\lambda$ for M-stack
(thick solid line) and H-stack (thick dashed line).  Asymptotics
of the localization length $l$ (thin straight lines), the short
wavelength asymptotic (thin dotted line, (\ref{shortWave}),
and the long wavelength asymptotics---thin solid line for
the M-stack (\ref{lloc}) and  a thin dashed line for the H-stack (\ref{asympt-H}).} \label{Fig2}
\end{figure}

%%%%%%%%%%%%%%%%%%%%%%%%%%%%%%%%%%%%%%%%%%%%%%%%%%%%%%%%%%%%%%%%%%%%%
We first consider stacks having refractive index and thickness disorder, with $Q_{\nu }=0.25$ and $Q_{d}=0.2$. Shown in Fig. \ref{Fig2} are transmission
spectra for a M-stack of $N=10^{5}$ layers and a H-stack of length $N=10^{3}$. There are two major differences between the results for these two
types of samples: first, in the localized regime ($N\gg l_{N}$), the
transmission length of the M-stack exceeds or coincides with that of the
H-stack; second, in the long wavelength region, the plot of the transmission
length of the M-stack exhibits a  pronounced bend, or kink, in the
interval $\lambda \in [ 10^{2},10^{3}]$, while there is no such
feature in the H-stack results. These two types of behaviour are discussed
in more detail below.

%%%%%%%%%%%%%%%%%%%%%%%%%%%%%%%%%%%%%%%%%%%%%%%%%%%%%%%%%
%%%%%%%%%%%%%%%%%%%%%%%%%%%%%%%%%%%%%%%%%%%%%%%%%%%%%%%

%%%%%%%%%%%%%%%%%%%%%%%%%%%%%%%%%%%%%%%%%%%%%%%%%%%%%%%%%%%%%%%%%%%%%

\subsubsection{Mixed stacks}

\label{subsubsec:mixeds}
%%%%%%%%%%%%%%%%%%%%%%%%%%%%%%%%%%%%%%%%%%%%%%%%%%%%%%%%%%%%%%%%%%%%%
The weak scattering approximation (WSA) of Eq. (\ref{translength-1})
is an excellent method by which to calculate the transmission length for M-stacks.
This is seen in
Fig. \ref{Fig2} where the curves obtained by numerical simulations and by the WSA
are indistinguishable (solid line). The characteristic wavelengths (\ref%
{wavelengths-3}) of this mixed stack are $\lambda _{1}\approx 148$ and $%
\lambda _{2}\approx 839$. Therefore, the transmission length describes the
localization properties of a random sample in the region $\lambda \lesssim
148$, whereas longer wavelengths, $\lambda \gtrsim 839$, correspond to the
ballistic regime. The crossover from the localized to the ballistic regime
demonstrates the  kink-type behaviour that occurs within the
region $\lambda _{1}\lesssim \lambda \lesssim \lambda _{2}$. The short and
long wavelength behaviour of the transmission length is also in excellent
agreement with the calculated asymptotics in both regimes.

%%%%%%%%%%%%%%%%%%%%%%%%%%%%%%%%%%%%%%%%%%%%%%%%%%%%
\begin{figure}[h]
\rotatebox{0}{\scalebox{0.4}{\includegraphics{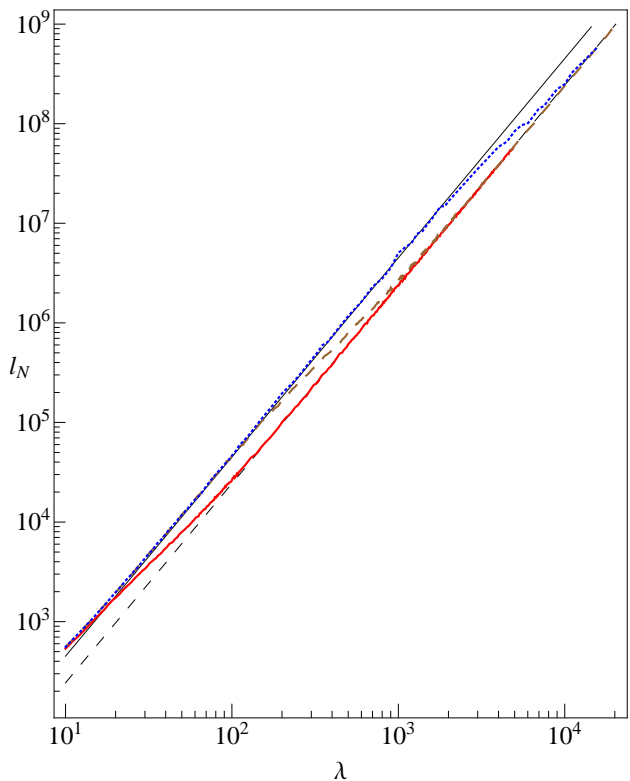}}}
\caption{ (Color online) Transmission length $l_{N}$ vs  $\lambda $ for a M-stack
of $N=10^{3}$ (thick solid line), $10^{5}$ (thick dashed line) and $
10^{7}$ (thick dotted line) layers showing both numerical
simulations and the WSA theory. The long wave asymptotics for the
localization length (\ref{lloc}) and the ballistic length
(\ref{bbal}) are shown respectively in the thin solid and dashed
lines respectively.} \label{Fig3}
\end{figure}
%%%%%%%%%%%%%%%%%%%%%%%%%%%%%%%%%%%%%%%%%%%%%%%%%%%%%%%%%%%%%

To analyze the long wavelength region ($\lambda\gtrsim 10$) more carefully,
we plot in Fig.~\ref{Fig3} the transmission lengths of M-stacks of three
different sizes, $N=10^{3},\ 10^{5},\ \text{and}\ 10^{7}.$ In all cases,
there is excellent agreement between the simulations and the WSA
predictions. The characteristic wavelengths for $N=10^{3}$ are $\lambda
_{1}=14.8$ and $\lambda_{2}=83.9$, while for $N=10^{7}$ they are $\lambda
_{1}=1480$ and $\lambda_{2}=8390$, and we see that the observed crossover
regions are bounded exactly by these characteristic wavelengths in all three
cases.

\begin{figure}[h]
\rotatebox{0}{\scalebox{0.51}{\includegraphics{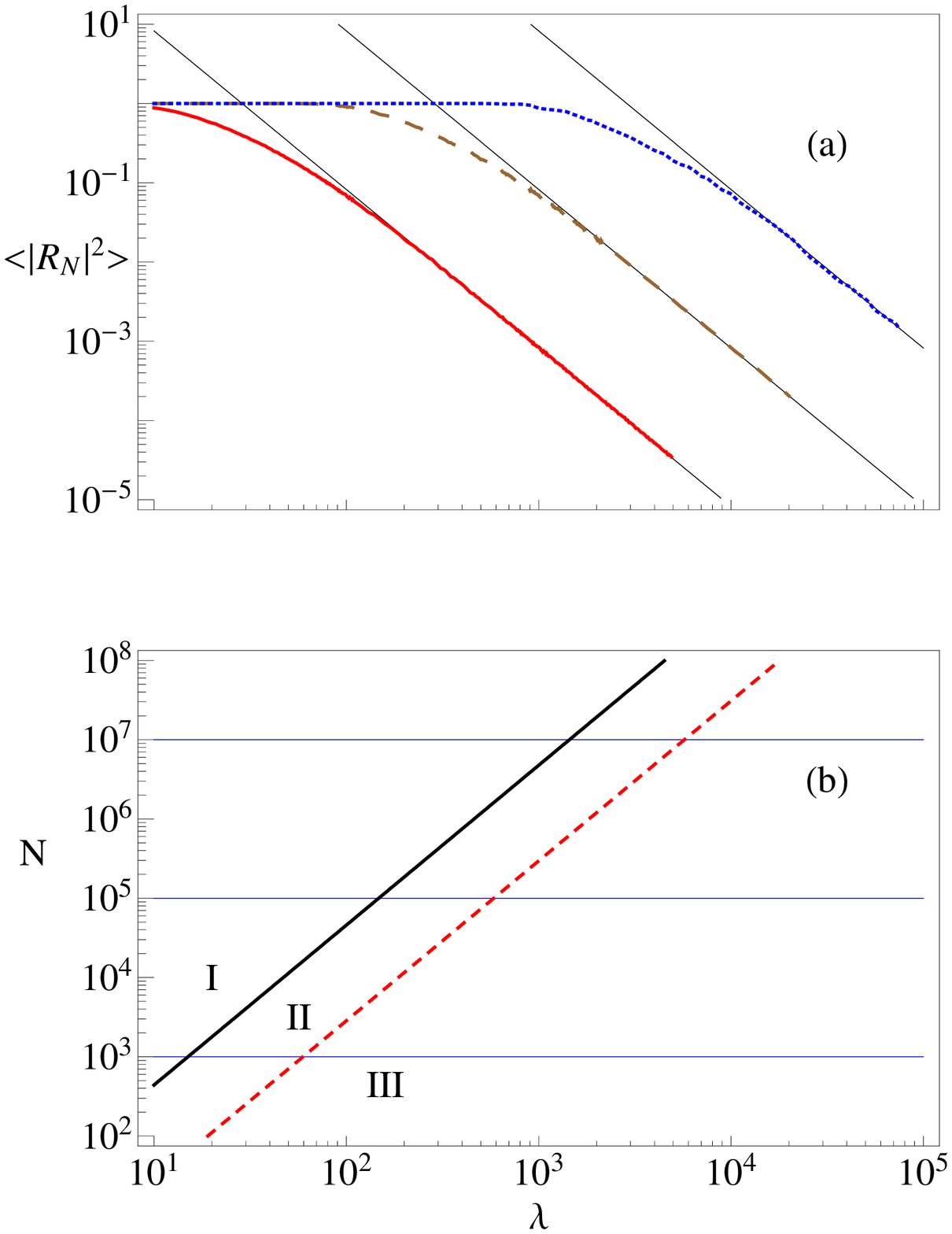}}}
\caption{ (Color online) (a) Average reflectance  for M-stacks of length
$N=10^3$ (solid line), $10^{5}$ (dashed line)
and $10^{7}$ (dotted line) layers (numerical simulation and WSA).
Long wave asymptotic for the average reflectance
 for the same stacks (thin solid lines).
\newline (b) Phase diagram of M-stacks. The thick solid line corresponds
to a stack size equal to the localization length. The dashed line
corresponds to a stack size equal to the crossover length. The
localized, crossover and ballistic regimes occur in regions I, II,
and III respectively.} \label{Fig4}
\end{figure}

To confirm the ballistic nature of the transmission in the region $\lambda
>\lambda _{2}$, we plot in Fig.~\ref{Fig4}(a) the logarithm of the mean
value of the reflectance for the same three stack sizes as a function of the logarithm
of the wavelength. In all cases, the plots exhibit a linear dependence $\ln
\langle |R_{N}|^{2}\rangle =\text{const}+2\ln \lambda $ in the ballistic
regime, which is bounded from below by the crossover wavelength $\lambda _{2}$. The straight lines are calculated from Eq. (\ref{totalRefl}) and confirm that
the reflection coefficient in the mixed stack is proportional to the stack
length. Within the localized region $\lambda \lesssim \lambda _{1}$, the
reflection coefficient is close to unity in all three cases.

The behaviour of the transmission length is illustrated by the phase diagram
in the $(\lambda ,N)$ plane shown in Fig.~\ref{Fig4}(b). The two slanted lines
$N=l(\lambda )$ and $N=\bar{l}(\lambda )$ separate the plane into three
parts corresponding to the localization (I), the crossover region (II) and
the ballistic region (III). The intersections of these lines with the
horizontal lines $N=10^{3}$, $N=10^{5}$, and $N=10^{7}$ define the
characteristic wavelengths $\lambda _{1}$ and $\lambda _{2}$ for the three
stack sizes considered here. It is easy to see that these wavelengths,
determined with the aid of the phase diagram, perfectly bound the crossover
regions in both Fig.~\ref{Fig3} and Fig.~\ref{Fig4}(a).

%%%%%%%%%%%%%%%%%%%%%%%%%%%%%%%%%%%%%%%%%%%%%%%%%%%%%%%%%%%%%%%
%%%%%%%%%%%%%%%%%%%%%%%%%%%%%%%%%%%%%%%%%%%%%%%%%%%%%%%%%%%%5
%%%%%%%%%%%%%%%%%%%%%%%%%%%%%%%%%%%%%%%%%%%%%%%%%%%%%%%%%%%%%%%%%%%%%

Until now, we have dealt only with the transmission length $l_N(\lambda)$, which was defined
through an average value.  However, more detailed information can be obtained from the transmission
length $\tilde{l}_N(\lambda)$ for a single realization, defined by the equation

%[single-real]
\begin{equation}
\frac{1}{\tilde{l}_{N}}=-\frac{\ln \ \left\vert T_{N}\right\vert }{N}.
\label{single-real}
\end{equation}

%%%%%%%%%%%%%%%%%%%%%%%%%%%%%%%%%%%%%%%%%%%%%%%%%%%%%%%%%%%%%%%
\begin{figure}[h]
\rotatebox{0}{\scalebox{0.40}{\includegraphics{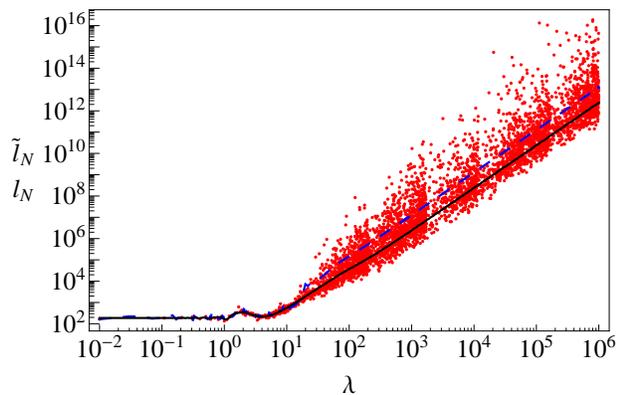}}}
\caption{(Color online) Transmission lengths $l_N$ (solid black line) and the transmission length for a single realization $\tilde{l}_N$ (dashed blue line) vs $\lambda$ for a M-stack with
$Q_{\nu}=0.25$, $Q_d=0.2$ and $N=10^4$ layers. Each separate point corresponds to a particular
wavelength with its own realization of a random stack.}
\label{Fig5New}
\end{figure}

In the localized regime, for a sufficiently long ({\em i.e.}, $N\gg l$)  M-stack, the transmission length for a single realization $\tilde{l}_{N}(\lambda)$ is  practically non-random and coincides with $l_N(\lambda)$, while in the ballistic region  it fluctuates.  The data displayed in Fig. \ref{Fig5New}) enables us to estimate the difference between
the transmission length $l_N(\lambda)$ (solid line) and the transmission length $\tilde{l}_N(\lambda)$
for a single randomly chosen realization (dashed line), and the scale of the corresponding fluctuations. Both curves are smooth, coincide in the localized region, and differ noticeably in the ballistic regime. The separate discrete points in Fig. \ref{Fig5New}  present the values of the transmission length $\tilde{l}_N(\lambda)$  calculated for different randomly chosen realizations. It is evident
that fluctuations in the ballistic region become more pronounced with increasing wavelength.

%, as is evident from  Fig.~\ref{Fig5New}. Here, for each wavelength $\lambda$, we chose a particular %realization and calculate the transmission length based on this realization at the given wavelength %$\lambda$. The resulting graph  $\tilde{l}_{N}(\lambda)$ displays strong fluctuations despite the  %smoothness (at least in the long wave region) of the graph  $\tilde{l}_{N}(\lambda)$, constructed for a %single realization over all wavelengths (see the insert to Fig.~\ref{Fig5New}).

\subsubsection{Homogeneous stacks}

\label{subsubsec:homoss}
%%%%%%%%%%%%%%%%%%%%%%%%%%%%%%%%%%%%%%%%%%%%%%%%%%%%%%%%%%%%%
%%%%%%%%%%%%%%%%%%%%%%%%%%%%%%%%%%%%%%%%%%%%%%%%%%%%%%%%%%%%%%%
\begin{figure}[h]
\rotatebox{0}{\scalebox{0.40}{\includegraphics{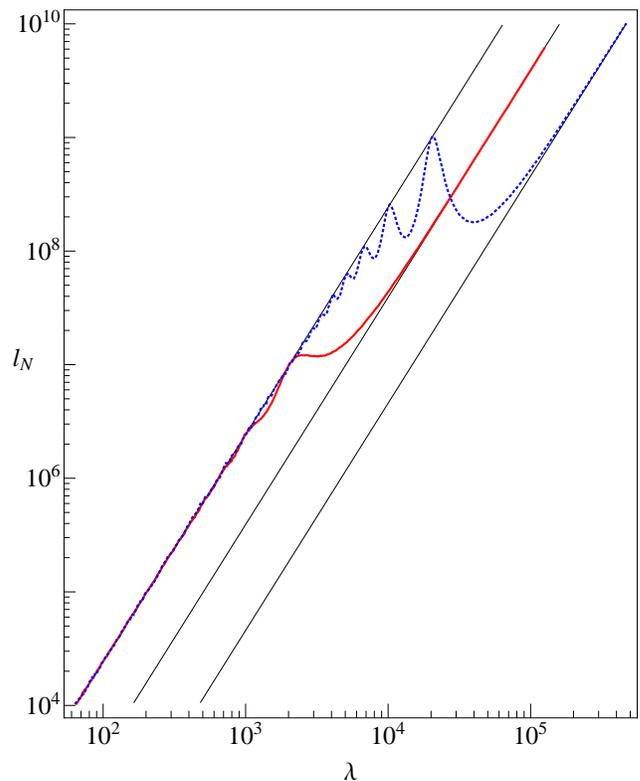}}}
\caption{ (Color online) Transmission length $l_{N}$ vs  $\protect\lambda $ for H-stacks
of $N=10^{3}$ (solid line), and $10^{4}$ (dotted line) layers
(numerical simulation and WSA). Long wave asymptotics for the
ballistic length in the near and far ballistic regions are plotted
in thin solid lines.}
\label{Fig5}
\end{figure}
%%%%%%%%%%%%%%%%%%%%%%%%%%%%%%%%%%%%%%%%%%%%%%%%%%%%%%%%%

The absence of any kink in the H-stack transmission length spectrum in Fig.~%
\ref{Fig2} follows from Sec.~\ref{subsec:hom} in which it was shown that the
crossover to the far ballistic regime occurs at the wavelength $\lambda
_{2}(N)$ (\ref{crossover-H-2}). For $N=10^{3}$, this is of the order of $10^{4}$ and so the kink does not appear.

In order to study the crossover, we plot in Fig.~\ref{Fig5} the
transmission lengths of H-stacks with  $N=10^{3}$ and $10^{4}$ over the
wavelength range extended up to $\lambda \sim 10^{6}$ . As for the M-stack, the simulation results for H-stacks cannot be distinguished from those of
the WSA (\ref{transhom}). The transition from the localized to the near
ballistic regime occurs without any change in the analytical dependence of
transmission length, in complete agreement with the results of Sec.~\ref%
{subsec:hom}. The crossover from the near to the far ballistic regime is
accompanied by a change in the analytical dependence that occurs at $%
\lambda =\lambda _{2}(N)$, which for these stacks is of the order of $10^{4}$
and $10^{5}$ respectively. The crossover is accompanied by prominent
oscillations described by Eq. (\ref{bal-H-6}). Finally, we note that the
vertical displacement between the moderately long and extremely long wavelength
ballistic asymptotes does not depend on wavelength, but grows with the size of
the stack, according to the law

%[shift]
\begin{equation}
\ln \frac{b_{n}}{b_{f}}=\ln \frac{NQ_{\nu }^{2}}{12},  \label{shift}
\end{equation}%
which stems from Eq. (\ref{bal-H-5}).

\begin{figure}[h]
\rotatebox{0}{\scalebox{0.51}{\includegraphics{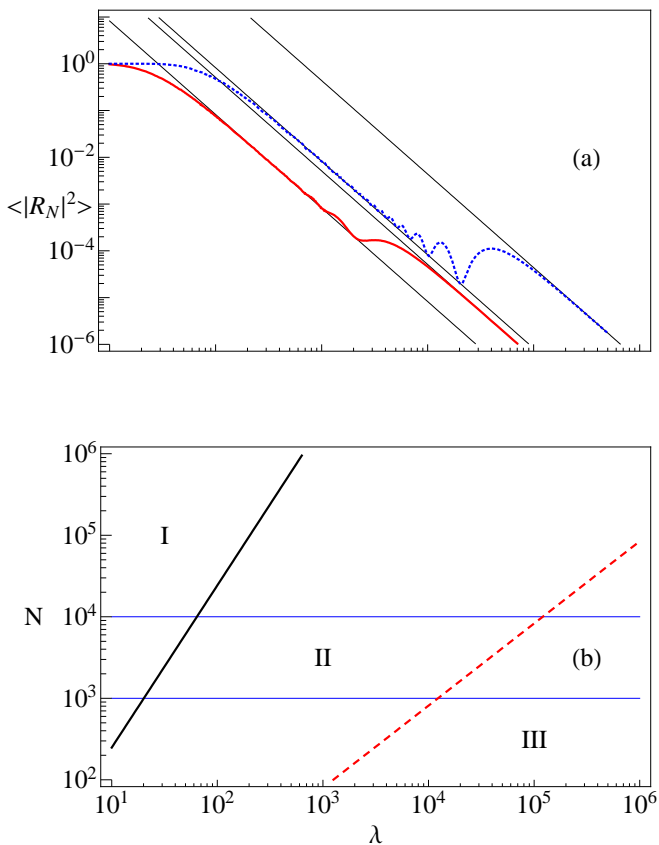}}}
\caption{ (Color online) (a) Average reflectance from  H-stacks of $N=10^3$ (solid
line), and $10^{4}$ (dotted line) layers (numerical simulation and
WSA). Long wave asymptotics of the ballistic length in the near and
far ballistic regions are plotted by thin solid lines. (b) Phase diagram for a  H-stack. The thick solid line
corresponds to where the stack size equals the localization length.
The dashed line corresponds to where the stack size equals the
crossover length, while the localized, near ballistic and far
ballistic regimes correspond to  regions (I), (II), and (III)
respectively.} \label{Fig6}
\end{figure}

To study the ballistic transmission regime in the region $\lambda >\lambda
_{1}$ more closely, we plot in the upper panel of Fig.~\ref{Fig6}(a) (on a
logarithmic scale) the mean value of the reflection coefficient for the same
two stack sizes. That part of each plot which presents the ballistic
propagation comprises two linear asymptotes of the form $\ln \langle
|R_{N}|^{2}\rangle =\text{const}+2\ln \lambda $, corresponding to different
values of the constant, one applicable in the near ballistic region $\lambda
_{1}(N)\lesssim \lambda \lesssim \lambda _{2}(N)$ and the other in the far
ballistic region $\lambda _{2}(N)\lesssim \lambda $. The difference between the
values of these two constants corresponds precisely with the right hand side of
Eq. (\ref{shift}), with the far long wavelength asymptotic given by the
second term in Eq. (\ref{bal-H-6}). Within the localized regime $\lambda
\lesssim \lambda _{1}(N),$ the reflectance is almost unity in all three
cases.

%%%%%%%%%%%%%%%%%%%%%%%%%%%%%%%%%%%%%%%%%%%%%%%%%%%%%%
%%%%%%%%%%%%%%%%%%%%%%%%%%%%%%%%%%%%%%%%%%%%%%%%%%%%%%%%%%%%%%%%

By analogy with M-stacks, the behaviour of the transmission length in the $%
(\lambda ,N)$ plane is illustrated by the phase diagram displayed in the
lower panel of Fig.~\ref{Fig6}. Two slanted lines $N=l(\lambda )$ and $N=\bar{%
l}(\lambda )$ divide the plane into three parts corresponding to the
localization regime (I), the crossover regime (II) and the ballistic regime
(III). The wavelength, at which these  lines intersect the horizontal
lines corresponding to the stack lengths,  $N=10^{3}$, and $N=10^{4}$,
defines the characteristic wavelengths (\ref{crossover-H-3}) for each stack
size. From Fig.~\ref{Fig6}(b), it follows that the separation between them grows
with increasing size in accordance with Eq. (\ref{shift}). It is easy to see
that the wavelengths $\lambda _{1,2}$ determined from the phase diagram
perfectly bound the near ballistic region in both Fig.~\ref{Fig5} and Fig.~%
\ref{Fig6}(a).

%As mentioned in Section \ref{subsec:hom}, the transmission length for
%any single realization in the far ballistic region becomes non random.  To
%confirm this observation we shown in  Fig.~\ref{Fig7} that although $%
%\tilde{l}_{N}$ exhibits fluctuations in the near ballistic region, it is a
%fairly smooth function of  wavelength in the far ballistic regime.

%%%%%%%%%%%%%%%%%%%%%%%%%%%%%%%%%%%%%%%%%%%%%%%%%%%%%%
\begin{figure}[h]
\rotatebox{0}{\scalebox{0.40}{\includegraphics{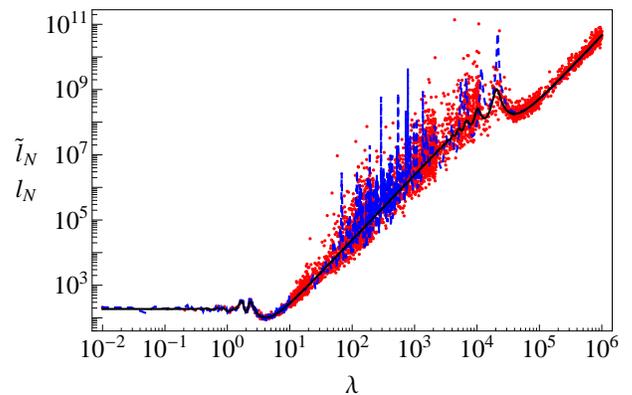}}}
\caption{(Color online) Transmission lengths $l_N$ (solid black line) and the transmission length for a single realization $\tilde{l}_N$ (dashed blue line) vs $\lambda$ for a H-stack with
$Q_{\nu}=0.25$, $Q_d=0.2$ and $N=10^4$ layers. Each separate point corresponds to a particular
wavelength with its own realization of a random stack.}
\label{Fig8New}
\end{figure}
%%%%%%%%%%%%%%%%%%%%%%%%%%%%%%%%%%%%%%%%%%%%%%%%%%%%%%%%%%

%%%%%%%%%%%%%%%%%%%%%%%%%%%%%%%%%%%%%%%%%%%%%%%%%%%%%%
%\begin{figure}[h]
%\rotatebox{0}{\scalebox{0.4}{\includegraphics{Fig7.eps}}}
%\caption{Transmission length for  a single realization
%$\tilde{l}_N$ of a H-stack of  length $N=10^3$ and with $Q_{\nu}=0.25$, $Q_d=0$}
%\label{Fig7}
%\end{figure}
%%%%%%%%%%%%%%%%%%%%%%%%%%%%%%%%%%%%%%%%%%%%%%%%%%%%%%%%%%%%%%%%

%We now consider the transmission length of a H-stack computed with  only a single realization, but for %which  each wavelength is computed using a different (single) realization. As is evident from the main %part of Fig. \ref{Fig8New}), we see that for extremely long stacks the behaviour  becomes practically %non-random, not only within the localized region as for a M-stack, but also in the far ballistic region %because of the self-averaging nature of the  effective dielectric constant (\ref{effective-4}). %However, these fluctuations are moderated since they should vanish in the limit as $N\to\infty$.  In %contrast, the graph in the inset to Fig. \ref{Fig8New}), which plots $\tilde{l}_{N}(\lambda)$ with each %wavelength computed using the same (single) realization, is smooth.

We now consider the transmission length for a  H-stack computed using a single realization. For extremely long stacks ($N \to \infty$), the transmission length becomes practically non-random, not only within the localized region (as is the case for M-stacks) but also in the far ballistic region because of the self-averaging nature of the  effective dielectric constant (\ref{effective-4}).
For less long stacks, however,  $\tilde{l}_N(\lambda)$ also fluctuates in the far ballistic region.
To demonstrate this, we have plotted in Fig. \ref{Fig8New} the transmission length $l_N$ (solid line) and the transmission length $\tilde{l}_N(\lambda)$ for a single randomly chosen realization (dashed line). It is evident that, in contrast to the results for the M-stack, the H-stack single realization transmission length  in the near ballistic region is a complicated and irregular function, similar to the well known ``magneto fingerprints'' of magneto-conductance of a disordered sample in the weak localization regime \cite{Altshu}. In the far ballistic region, these fluctuations are moderated  since they  vanish in the limit as $N\to\infty$.  To support this statement, we display in Fig. \ref{Fig8New} the set of separate discrete points, each
of them presenting $\tilde{l}_N(\lambda)$ calculated for a different randomly chosen realization.

In summary, we note that the results of Sec. \ref{subsec:refthick} show
excellent agreement between the numerical simulations and the analytical
predictions of the weak scattering approximation of Sec.~\ref{sec:anal}.
Moreover, even for the case of an H-stack of  length $N=10^{3}$ and
strong disorder, $Q_{\nu }=0.9$ and $Q_{d}=0.2$, the results of direct
simulation and those of the WSA analysis coincide completely in the long
wavelength region and differ by only a few percent in the short wave region where the scattering is certainly not weak. This is because the
perturbation approach based on Eqs. (\ref{rec3}), (\ref{rec4}) is related
not to the calculated quantities but to the equations they satisfy.

\subsection{Refractive-index disorder}

\label{subsec:ref} %%%%%%%%%%%%%%%%%%%%%%%%%%%%%%%%%%%%%%%%%%%%%%%%%%

\subsubsection{Suppression of localization}

\label{subsubsec:suppress}
%%%%%%%%%%%%%%%%%%%%%%%%%%%%%%%%%%%%%%%%%%%%%%%%%%%%
Here, we present  results for stacks with only refractive index disorder
(RID), with $Q_{\nu }=0.25$. For H-stacks, the transmission length
demonstrates qualitatively and quantitatively the same behaviour as was
observed in the presence of both refractive index and thickness disorder.
Corresponding formulae for the transmission, localization, and ballistic
lengths can be obtained from the general case by taking the limit as $%
Q_{d}\rightarrow 0$.

\begin{figure}[h]
\rotatebox{0}{\scalebox{0.40}{\includegraphics{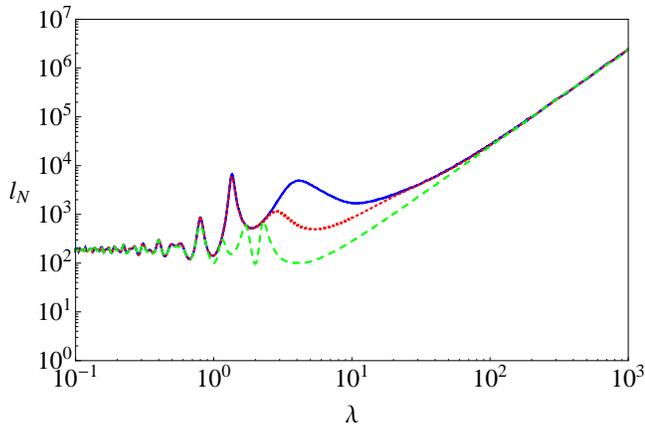}}}
\caption{(Color online) Transmission length $l_N$ vs wavelength $\lambda$ for
stacks of  $N=10^3$ layers. M-stack: numerical simulation (solid
line), WSA (dotted middle line). H-stack: numerical simulation and WSA
(both dashed lines).} \label{Fig8}
\end{figure}

In the case of M-stacks, however, the situation changes markedly. While in
the short wavelength region and in the ballistic regime the numerical
results are still in a excellent agreement with the predictions of the weak
scattering approximation, the WSA fails in the long wavelength part of the
localization regime. This discrepancy manifests itself also for short
stacks, in which the localization regime in the long wave region is absent.
In Fig.~\ref{Fig8} we present the transmission length spectrum for a
M-stack of $N=10^{3}$ layers,  showing that, for $\lambda \approx 5$, the
numerical results for the transmission lengths of M-stacks differ by an order of magnitude from those
predicted by the WSA analysis, and also those observed for the corresponding H-stacks.

%%%%%%%%%%%%%%%%%%%%%%%%%%%%%%%%%%%%%%%%%%%%%%%%%%%%%%
%%%%%%%%%%%%%%%%%%%%%%%%%%%%%%%%%%%%%%%%%%%%%%%%%%%%%%%%%%

\begin{figure}[h]
\rotatebox{0}{\scalebox{0.40}{\includegraphics{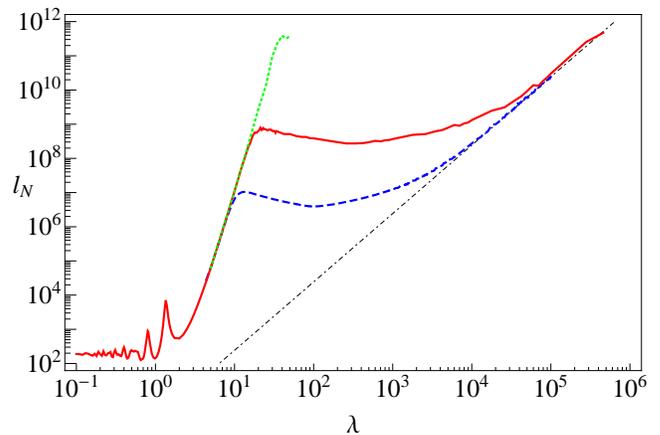}}}
\caption{(Color online) Transmission length $l_{N}$ vs wavelength $\lambda $ for an
M-stack with $Q_{\nu}=0.25$, $Q_{d}=0$ and $N=10^{7}$ (dashed line),
$N=10^{9}$ (solid line), and $N=10^{12}$ (dotted line) layers.}
\label{Fig9}
\end{figure}

For longer stacks, the differences in the transmission length spectra
exhibited by M-stacks and H-stacks becomes much more pronounced. In Fig.~\ref%
{Fig9}, we plot  transmission length spectra for different values of the
M-stack length with $N=10^{7}$ (using $10^{3}$ realizations), $10^{9}$
(using $10^{3}$ realizations) and $10^{12}$ layers (using only a single
realization), with the dashed-dotted straight line in Fig.~\ref{Fig9}
showing the long wavelength ballistic asymptote (\ref{bbal}). In the
moderately long wavelength region corresponding to the localization regime,
the transmission length $l_{N}\ll N$ coincides with the localization length $%
l$. It exceeds the H-stack localization length by a few orders of magnitude
and is characterized by a completely different wavelength dependence. This
substantial suppression of localization was revealed in our earlier paper
\cite{we} where the localization length $l_{N}$ was reported to be
proportional to $\lambda ^{6}$, in contrast to the classical $\lambda ^{2}$
dependence (\ref{lloc}) that is observed for H-stacks, and which is also
valid for M-stacks with both refractive index and thickness disorder. The
reason for this  difference is the lack of phase accumulation caused by the
phase cancelation in alternating layers of equal thicknesses.

%%%%%%%%%%%%%%%%%%%%%%%%%%%%%%%%%%%%%%%%%%%%%%%%%%%%%%
%%%%%%%%%%%%%%%%%%%%%%%%%%%%%%%%%%%%%%%%%%%%%%%%%%%%%%%%%%%%%

To study this behaviour in more detail, we generate a least squares fit $%
l_{N}=A\lambda ^{p}$ to the transmission length data. Respectively, for $N=%
10^{7}$, $10^{9}$, and $10^{12}$ layers, the best fits are $l_{N}=4\lambda
^{6.25}$, $l_{N}=0.43\lambda ^{7.38}$ and $l_{N}=0.01\lambda ^{8.78}$,
indicating that the asymptotic form for the localization length differs from
a pure power law, and perhaps is described by a non-analytic dependence.

%%%%%%%%%%%%%%%%%%%%%%%%%%%%%%%%%%%%%%%%%%%%%%%%%%%%%%%%%%%%%%%%
%\begin{figure}[h]
%\rotatebox{0}{\scalebox{0.40}{\includegraphics{Fig10.eps}}}
%\caption{Transmission length $l_N$ vs $\lambda$ for an M-stack with
%$Q_{\nu}=0.25$, $Q_d=0$ and $N=10^4$ layers (solid line) and the
%transmission length for a realization $\tilde{l}_N$ for the same
%values $Q_{\nu}, Q_d$ (dashed line). Each wavelength corresponds to
%a different realization of the refractive index.} \label{Fig10}
%\end{figure}
%%%%%%%%%%%%%%%%%%%%%%%%%%%%%%%%%%%%%%%%%%%%%%%%%%%%%%%%%%%%%%%%

In the crossover part of the long-wave region where $l_{N}\approx N$, the
transmission length of M-stacks also differs essentially from that for
H-stacks. Moreover, the width of the crossover region on the $l_{N}-$%
axis remains of the same order of magnitude, although on the $\lambda -$axis
it grows for $N=10^{9}$ to four orders of magnitude, being much wider for
longer stacks. In the long wavelength region corresponding to the
ballistic regime, the transmission length of the RID M-stack coincides with
that of RID H-stack.

%As was mentioned above, the transmission length $\widetilde{l}(N)$ for  a single
%realization approaches its non-random limit $l$ as $N \to \infty$.
%This means that in the localization region, the dependence $\widetilde{l}(N)$
%should be rather smooth because of self-averaging, while outside this region
%fluctuations must occur. To check this, we plot in Fig.\ref{Fig10}  $%
%\widetilde{l}(N)$ and the transmission length $l_{N}$ as functions of
%wavelength $\lambda $ for a RID M-stack of $N=10^{4}$ layers. We note that
%in the localized part of the spectrum, $\lambda <5$, the ensemble averaged
%and single realization curves are close, revealing that the self-averaging
%property of $\widetilde{l}(N)$ holds for mixed media. Within the crossover
%region, however, the dependence of $\widetilde{l}$ on $\lambda $
%demonstrates prominent fluctuations.

%%%%%%%%%%%%%%%%%%%%%%%%%%%%%%%%%%%%%%%%%%%%%%%%%%%%%%%%%%%

\subsubsection{Transmission resonances}

\label{subsubsec:transres}
%%%%%%%%%%%%%%%%%%%%%%%%%%%%%%%%%%%%%%%%%%%%%%%%%%%%%%%%%%%%%%%%
An important signature of the localization regime is the presence of
transmission resonances (see, for example, Refs. \cite%
{Lifshits,Azbel,Bliokh-2}), which appears in sufficiently long, open systems
and which are a ``fingerprint''  of a given
realization of disorder. These resonances are responsible for the difference
between two quantities that characterize the transmission, namely $\langle
\ln |T|^{2}\rangle $ and $\ln \langle |T|^{2}\rangle $. The former reflects
the properties of a typical realization, while the main contribution to the
latter is generated by a small number of almost transparent realizations
associated with the transmission resonances.

%%%%%%%%%%%%%%%%%%%%%%%%%%%%%%%%%%%%%%%%%%%%%%%%%%%%%%%%%%%
\begin{figure}[h]
\rotatebox{0}{\scalebox{0.40}{\includegraphics{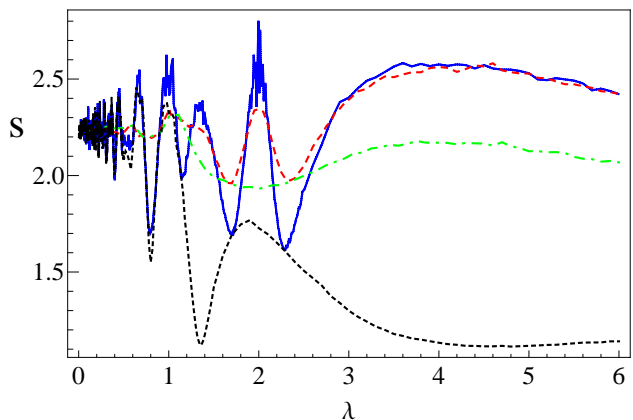}}}
\caption{(Color online) Ratio $s(\lambda)$ vs. wavelength $\lambda$ for $Q_{\protect\nu}=0.25$
and the stack length $N=10^3$. Solid and dashed curves are for the RID H-stack
and H-stack with $Q_d =0.2$, respectively. The middle dashed-dotted curve is
for an M-stack with $Q_d =0.25$, and the bottom dotted line is for a RID M-stack.}
\label{Fig11}
\end{figure}
%%%%%%%%%%%%%%%%%%%%%%%%%%%%%%%%%%%%%%%%%%%%%%%%%%%%%%%%%%%%%%%%

The natural characteristic of the transmission resonances is the ratio of
the two quantities mentioned above:

%[ratio]
\begin{equation}
s=\frac{\langle \ln |T|^{2}\rangle }{\ln \langle |T|^{2}\rangle }.  \notag
\label{ratio}
\end{equation}%
In the absence of resonances, this value is close to unity, while in the
localization regime $s>1$. In particular, in the high-energy part of the
spectrum of a disordered system with Gaussian white-noise potential, this
ratio takes the value $4$ \cite{LGP}.

%%%%%%%%%%%%%%%%%%%%%%%%%%%%%%%%%%%%%%%%%%%%%%%%%%%%%%%%
\begin{figure}[h]
\rotatebox{0}{\scalebox{0.40}{\includegraphics{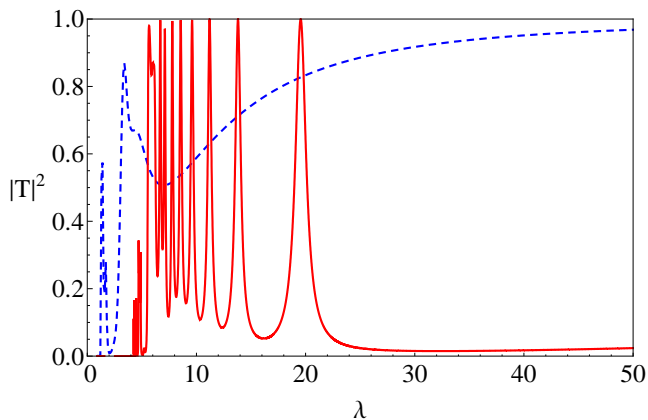}}}
\caption{ (Color online) Single realization transmittance $|T|^{2}$ vs wavelength
$\lambda$ for RID M-stacks with
$Q_\nu =0.25$ and $Q_{d}=0$ for $N=10^{5}$ layers (solid line)
and $N=10^{3}$ layers (dotted line).}
\label{Fig12}
\end{figure}
%%%%%%%%%%%%%%%%%%%%%%%%%%%%%%%%%%%%%%%%%%%%%%%%%%%%%%%%%%%%
In Fig.~\ref{Fig11}, we plot the ratio $s(\lambda )$ as a function of the
wavelength for RID M- and H-stacks and for the corresponding stacks with
thickness disorder. In all cases, the stack length is $N=10^{3}$ and it is
evident that for the RID M-stack $s(\lambda )\approx 1$, {\em i.e.}. the stack
length is too short for the localization regime to be realized. In other
three cases, however, $s(\lambda )\gtrsim 2$, which means that the
localization takes place even in a comparatively short stack.

%%%%%%%%%%%%%%%%%%%%%%%%%%%%%%%%%%%%%%%%%%%%%%%%%%%%%%%%%%%%%%
\begin{figure}[h]
\rotatebox{0}{\scalebox{0.40}{\includegraphics{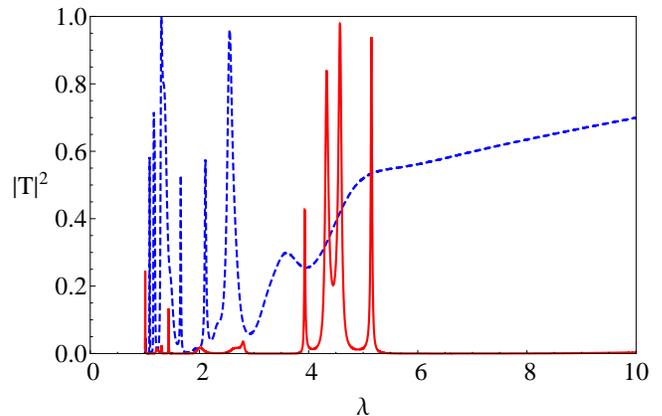}}}
\caption{ (Color online) Single realization transmittance $|T|^2$ vs
$\lambda$ for M-stack of $N=10^3$
layers with $Q_\nu=0.25$. Solid line corresponds to an M-stack with $Q_d=0.2$,
and the dashed line to  M-stack with no thickness disorder, {\em i.e.}, $Q_d=0.0$.}
 \label{Fig13}
\end{figure}
%%%%%%%%%%%%%%%%%%%%%%%%%%%%%%%%%%%%%%%%%%%%%%%%%%%%%%%%%%%%%%%
Thus, there are two ways in which to introduce transmission resonances. The
first is to increase the length of the stack. Fig.~\ref{Fig12} displays
the RID M-stack transmittance $|T|^{2}$ for a single realization as a
function of $\lambda $ for two lengths: $N=10^{5}$ (solid line) and $N=10^{3}
$ (dotted line). It is readily seen that while there are no resonances in
the shorter stacks, they do appear for the longer sample. The second way to generate transmission resonances is
to introduce thickness disorder. To demonstrate this, we plot in Fig.~\ref%
{Fig13} the transmittance of a single M-stack with both thickness and
refractive index disorder. It clearly shows that while the RID M-stack is
too short to exhibit transmission resonances at $\lambda >3$, resonances do emerge at longer wavelengths for the M-stack with thickness disorder.

%%%%%%%%%%%%%%%%%%%%%%%%%%%%%%%%%%%%%%%%%%%%%%%%%%%%%%%%

\subsubsection{Effects of the thickness disorder and
uncorrelated paring}

\label{subsubsec:diminish}
%%%%%%%%%%%%%%%%%%%%%%%%%%%%%%%%%%%%%%%%%%%%%%%%%%%%%%%%%

%%%%%%%%%%%%%%%%%%%%%%%%%%%%%%%%%%%%%%%%%%%%%%%%%%%%%%%%%
\begin{figure}[h]
\rotatebox{0}{\scalebox{0.40}{\includegraphics{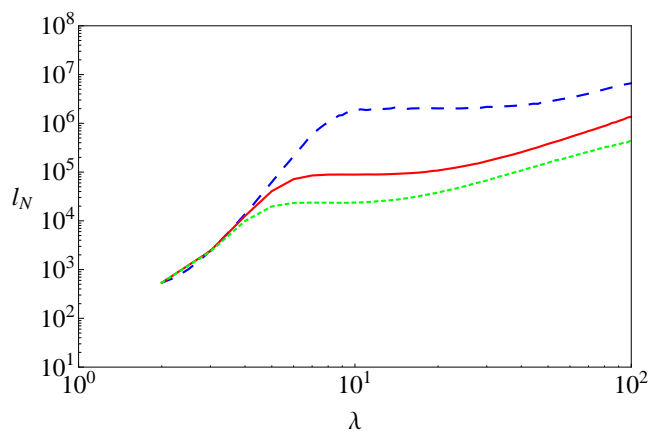}}}
\caption{(Color online) Localization length $l$ vs. wavelength $\protect\lambda$ for a
M-stack with $Q_\protect\protect\nu =0.25$ and $Q_d$=0.001, 0.005, and 0.01
(from top to bottom). }
\label{Fig14}
\end{figure}
%%%%%%%%%%%%%%%%%%%%%%%%%%%%%%%%%%%%%%%%%%%%%%%%%%%%%%%%%%%%%%%%
Here, we analyze the effect of thickness disorder on the $\lambda^6$-anomaly --- that is the $\lambda ^{6}$ dependence of the transmission length. In
Fig.~\ref{Fig14}, we plot the transmission length $l_{N}$ \ for an M-stack
with fixed refractive index disorder ($Q_{\nu }=0.25)$ for various values of
the thickness disorder. It is evident that the transmission length changes
from $l\propto \lambda ^{6}$ to the classical dependence $l\propto \lambda
^{2}$ as the thickness disorder increases from $Q_{d}=0.001$ (top curve) to $%
Q_{d}=0.01$ (bottom curve). In all cases, the number of layers $N=10^{8}$ is
longer than the transmission length,  guaranteeing  that Fig.~\ref{Fig14}
represents the genuine localization length $l$.

%%%%%%%%%%%%%%%%%%%%%%%%%%%%%%%%%%%%%%%%%%%%%%%%%%%%%%%%%%%
\begin{figure}[h]
\rotatebox{0}{\scalebox{0.40}{\includegraphics{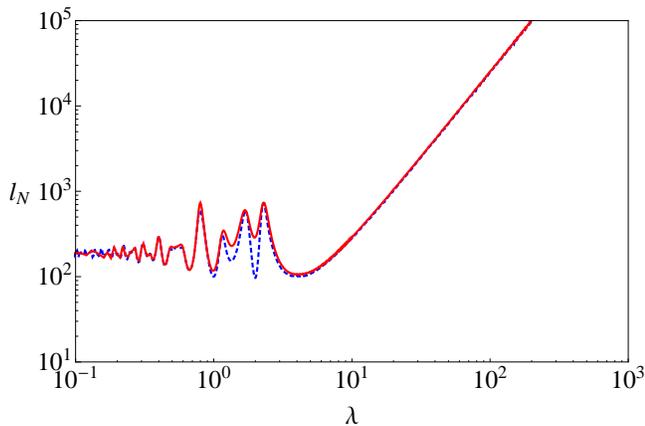}}}
\caption{(Color online) Transmission length $l_N$ vs. wavelength $\lambda $ for
the H-stack with R-layers (dashed line) and an M-stack (solid line)
in which each subsequent layer is chosen with equal probability to
be of R- or L-type. The stacks in both calculations are of the same
size, $N=10^4$.} \label{Fig15}
\end{figure}
%%%%%%%%%%%%%%%%%%%%%%%%%%%%%%%%%%%%%%%%%%%%%%%%%%%%%%%%%%%%%%
We have also found that the anomalous dependence $l\propto \lambda ^{6}$ (or
with higher power) is extremely sensitive to the alternation of left- and
right-handed layers. To demonstrate this, we consider an M-stack of length
$N=10^{4},$ in which each subsequent layer is chosen with equal probability
to be either right- or left-handed. Figure~\ref{Fig15} shows the
transmission length spectrum for this case, which is almost the same for
both the H-stacks and M-stacks. The only difference is a fairly modest
suppression of localization, which occurs within the wavelength interval $%
0.5<\lambda <2.5$. This result confirms that it is the additional
correlation between left-handed and right-handed layers in the alternating
stack which is responsible for the suppression of localization.

%%%%%%%%%%%%%%%%%%%%%%%%%%%%%%%%%%%%%%%%%%%%%%%

\subsection{Effects of losses}

\label{subsec:losses} %%%%%%%%%%%%%%%%%%%%%%%%%%%%%%%%%%%%%%%%
In this section, we study the transmission through layered media with absorption,
which is characteristic of real metamaterials. In this case, the
exponential decay of the field is due to both Anderson localization and
absorption \cite{FrePusYur,Asatrian2}, and in some limiting cases, it is
possible to distinguish between these contributions.

For an M-stack with weak fluctuations of the refractive index, weak
thickness disorder and  weak absorbtion, the WSA theory in the limits of short or long waves leads to
the well-known formula

%[att-l]
\begin{equation}
\frac{1}{l_{\mathrm{att}}}=\frac{1}{l_{N}}+\frac{1}{l_{\mathrm{abs}}},
\notag  \label{att-l}
\end{equation}%
where $l_{N}$ is the disorder-induced transmission length in the absence of
absorption, and the absorption length is

%[abs]
\begin{equation}
l_{\mathrm{abs}}=\frac{\lambda }{2\pi \sigma }.  \label{abs}
\end{equation}

For short wavelengths, $l_{N}^{-1}$ is a constant given by Eqs (\ref{shortWave}%
), (\ref{bbal}) and (\ref{lloc}), while for long wavelengths, in either the
localized or ballistic regimes, its contribution is proportional to $%
\lambda ^{-2}$ and thus is negligibly small in comparison with the contribution due to losses, which is always proportional to $\lambda ^{-1}$. Accordingly, at
both short and long wavelengths, the attenuation length coincides with the
absorption length (\ref{abs}), with disorder contributing significantly to
the attenuation length only in some intermediate wavelength region, provided
that the absorption is sufficiently small.

%%%%%%%%%%%%%%%%%%%%%%%%%%%%%%%%%%%%%%%%%%%%%%%%%%%%%%%%%%%%%%
\begin{figure}[h]
\rotatebox{0}{\scalebox{0.40}{\includegraphics{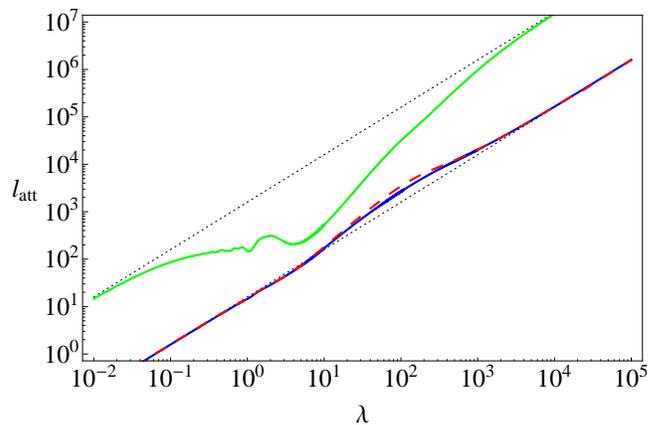}}}
\caption{(Color online) Attenuation length $l_{\mathrm{att}}$ and absorption length
$l_{\mathrm{abs}}$ vs. wavelength $\lambda$ for an M-stack with
disorder $Q_\nu=0.25$, $Q_d=0.2$ and  length $N=10^4$. The upper
solid line displays the (identical) simulation and WSA results for
$\sigma=10^{-4}$; the lower solid line presents numerical results
 while the dashed line displays WSA results for the same
absorption value. Absorption lengths (\ref{abs}) for both
$\sigma=10^{-4}$ and $\sigma=10^{-2}$  are shown by dotted straight
lines.} \label{Fig16}
\end{figure}
%%%%%%%%%%%%%%%%%%%%%%%%%%%%%%%%%%%%%%%%%%%%%%%%%%%%%%%%%%%

The results of the numerical calculations shown in Fig.~\ref{Fig16}
completely confirm the theoretical predictions presented above. For weak
absorption $\sigma =10^{-4}$, the direct simulation and WSA theory  give
exactly the same result (solid curve in Fig.~\ref{Fig16}). Over a reasonably
wide wavelength range, $10^{-1}\lesssim \lambda \lesssim 10^{3}$, disorder
contributes significantly to the attenuation. For such a stack, the
characteristic wavelengths are $\lambda _{1}(N)= 47$ and $\lambda
_{2}(N)= 265$, implying that the contribution of disorder is
significant in all regions, from the short wavelength part of the localized
regime to the long wavelength ballistic region. Due to the losses, however,
there are fewer oscillations evident in the transmission length spectrum
than in the lossless case of Fig.~\ref{Fig2}.

For stronger absorption, $\sigma =10^{-2}$, the wavelength range, over which
disorder contributes to the attenuation is reduced, as well as the relative
value of the contribution itself. The agreement between the numerical
simulations and the WSA calculations is reasonable.

%%%%%%%%%%%%%%%%%%%%%%%%%%%%%%%%%%%%%%%%%%%%%%%%%%%%%%%%%%%%

\section{Conclusions}

\label{sec:conc}
%%%%%%%%%%%%%%%%%%%%%%%%%%%%%%%%%%%%%%%%%%%%%%%%%%%%%%%%%%%%%%%

We have studied the transmission and localization of classical waves in
one-dimensional disordered structures composed of alternating layers of
left- and right-handed materials (M-stacks) and have compared this to the
transport in homogeneous structures composed of different layers of the same
material (H-stacks). For weakly scattering layers and general disorder,
where both refractive index and thickness of each layer is random, we have
developed an effective analytical approach, which has enabled us to
calculate the transmission length over a wide range of input parameters and
to describe transmission through M- and H-stacks in a unified way. All
theoretical predictions are in excellent agreement with the results of
direct numerical simulations.

There are remarkable distinctions between the transmission and localization
properties of the two types of stacks. When both types of disorder
(refractive index and layer thickness) are present, the transmission length
of a H-stack in the localized regime coincides with the reciprocal of the Lyapunov exponent, while for M-stacks these two quantities differ by a numerical prefactor. This is quite surprising and, to the best of our knowledge, it is  the
first time that a system where such a difference exists has been discovered.

It is shown that the stacks of M- and H-type manifest quite different
behaviour of the transmission length as a function of wavelength. This
difference is most pronounced in the long wavelength region. In the localized
regime, stacks of both types are strongly disordered and reflect the incident
wave almost entirely. In the ballistic regime, where stacks are almost
transparent, the transport properties of H-stacks and M-stacks are markedly
different. H-stacks, over the moderately long wavelength ballistic region,
and M-stacks, over the entire long wavelength region, are weakly scattering
disordered structures. In the extremely long wavelength ballistic region,
the H-stack becomes effectively uniform --- a regime which is absent for
M-stacks because of the intrinsic non-uniformity caused by the alternating
nature of the structure. The crossover regions between different regimes are
comparatively narrow.

The transmission length for a single realization is non-random in the
localized regime for both types of stacks. It  fluctuates strongly over the entire ballistic region for M-stacks and in the near ballistic regime for H-stacks. In the far long wave region for  H-stacks, the fluctuations apparent in the transmission length are moderate and decrease with increasing stack length.  In the case of M-stacks, the transition from the localized to the ballistic regime is accompanied by a change in the
wavelength dependence of the transmission length. In contrast, for H-stacks,
the corresponding change occurs in the vicinity of the transition from the near to the far ballistic regime.   Again, the crossover regions between the different regimes are comparatively narrow.

In M-stacks with only refractive-index disorder, Anderson localization is
substantially suppressed, and the localization length grows with increasing
wavelength much faster than the classical square law dependence. The
crossover region becomes significantly wider, and transmission resonances
occur in much longer stacks than in the corresponding H-stacks.

The effects of absorption on the one-dimensional transport and localization
have also been studied, both analytically and numerically. In particular, it
has been shown that the crossover region is particularly sensitive to
losses, so that even small absorption noticeably suppresses  the oscillations
of the transmission length in the frequency domain.

%%%%%%%%%%%%%%%%%%%%%%%%%%%%%%%%%%%%%%%%%%%%%%%%%%%%%%%%%%%%%%%%%%%%%%%%

\section{Acknowledgments}

\label{sec:ackn}
%%%%%%%%%%%%%%%%%%%%%%%%%%%%%%%%%%%%%%%%%%%%%%%%%%%%%%%%%%%%%%%%%%%%%%%%

This work was supported by the Australian Research Council through
the Discovery and Centres of Excellence programs, and also by the Israel
Science Foundation (Grant \# 944/05). We thank L. Pastur, K. Bliokh and Yu.
Bliokh for useful discussions.  We also acknowledge the provision of
computing facilities through NCI (National Computational Infrastructure) and Intersect in Australia.\\

%%%%%%%%%%%%%%%%%%%%%%%%%%%%%%%%%%%%%%%%%%%%%%%%%%%%%%%%%%%%%%%%%%%%%
\appendix*

\section{Uniform distribution of fluctuations}

\label{App} %%%%%%%%%%%%%%%%%%%%%%%%%%%%%%%%%%%%%%%%
If the fluctuations $\delta _{j}^{(\nu )}$ and $\delta _{j}^{(d)}$ are
uniformly distributed in the intervals $[-Q_{\nu },Q_{\nu }]$ and $%
[-Q_{d},Q_{d}]$ respectively, and absorption is the same in all slabs ($%
\sigma _{j}=\sigma $), the transmission, localization, and ballistic lengths
given by Eqs.~(\ref{FinalG3})-(\ref{l}) can be calculated explicitly in the
weak scattering approximation. The results are:

%[B1]
\begin{eqnarray}
\langle t^{2}\rangle &=&\frac{1}{8ikQ_{\nu }Q_{d}} \left[ Ei(i\Delta
_{+}n_{+})-Ei(i\Delta _{-}n_{+})\right.  \notag \\
&-&\left. Ei(i\Delta _{+}n_{-})+Ei(i\Delta _{-}n_{-})\right] ,  \label{A-1}
\end{eqnarray}
%B2
\begin{eqnarray}
\langle r\rangle &=&\frac{e^{i\Delta _{+}(1+i\sigma )}}{8ikQ_{d}} \left[%
\frac{\sin (\Delta _{+}Q_{\nu })}{\Delta _{+}Q_{\nu }}- \frac{\sin
(\Delta_{-}Q_{\nu }))}{\Delta _{-}Q_{\nu }}\right] +  \notag \\
&+&\frac{i\sigma }{2}(\langle t^{2}\rangle -1)+ \frac{1+i\sigma }{2}\langle
t^{2}\rangle ,  \label{A-2}
\end{eqnarray}

%[avlnt]
\begin{eqnarray}
\langle |r|^{2}\rangle &=&2k\sigma +\frac{Q_{\nu }^{2}}{6}- \frac{\sigma ^{2}%
}{4}(\mathrm{Re}\langle t^{2}\rangle -1)  \notag \\
&+&2\sigma \mathrm{Im}\langle r\rangle -2\mathrm{Re}(H_{1}+H_{2}).
\label{A-3}
\end{eqnarray}%
Here,

%[del],[npm],[H]
\begin{eqnarray*}
\Delta _{\pm } &=&2k(1\pm Q_{d}),  \label{A-4} \\
n_{\pm } &=&1\pm Q_{\nu }+i\sigma ,  \notag \\
H_{1} &=&\frac{e^{i\Delta _{+}(1+i\sigma )}}{8kQ_{\nu }Q_{d}\Delta _{+}^{2}}%
\left[ 1+i\Delta _{+}(1+i\sigma )\sin (\Delta _{+}Q_{\nu })-\right. ,  \notag
\\
&-&\left. \Delta _{+}Q_{\nu }\cos (\Delta _{+}Q_{\nu })\right]  \notag \\
&+&\frac{i(1+i\sigma )^{2}}{16kQ_{\nu }Q_{d}}[Ei(i\Delta
_{+}n_{-})-Ei(i\Delta _{+}n_{+})],  \notag
\end{eqnarray*}%
where $H_{2}$ is obtained from $H_{1}$ by the replacement of $\Delta _{+}$
by $\Delta _{-}$, and $Ei(z)$ is the  exponential integral  given by
%[IntExp]
\begin{equation*}
Ei(z)=-\int_{-z}^{\infty }\frac{e^{-t}}{t}dt.  \label{A-5}
\end{equation*}
%%%%%%%%%%%%%%%%%%%%%%%%%%%%%%%%%%%%%%%%%%


\begin{thebibliography}{99}
\bibitem{Veselago} V. G. Veselago, Sov. Phys. Usp. \textbf{10}, 509 (1968).

\bibitem{Pendry} J. B. Pendry, Phys. Rev. Lett. \textbf{85}, 3966 (2000).

\bibitem{Sok} P. Marcos, C.M. Soukoulis, \textit{Wave propagation from
electrons to photonic crystals and left-handed materials}, (Princeton
University Press, Princeton, 2008).

\bibitem{Bliokh} K. Yu. Bliokh, and Yu. P. Bliokh, Physics - Uspekhi \textbf{%
47}, 393 (2004) [Usp. Fiz. Nauk \textbf{174}, 439 (2004)].

\bibitem{classification} C. Caloz and T. Ito, Proceedings of IEEE \textbf{93}%
, 1744 (2005).

\bibitem{Shalaev} V. M. Shalaev, Nature Photonics \textbf{1}, 41 (2007).

\bibitem{Schurig} D. Schurig, J. Mock, B. Justice, S. Cummer, J. Pendry, A.
Starr, and D. Smith, Science \textbf{314}, 977 (2006).

\bibitem{emis} J. K$\ddot{a}$stel and M. Fleischhauer, Phys. Rev. A \textbf{%
71}, 011804(R) (2005).

\bibitem{inter} Y. Yang, J. Xu, H. Chen, and S. Zhu, Phys. Rev. Lett.
\textbf{100}, 043601 (2008).

\bibitem{defect} L. G. Wang, H. Chen, and S.-Y. Zhu, Phys. Rev. B \textbf{70}%
, 245102 (2004).

\bibitem{Gorkunov} M. V. Gorkunov, S. A. Gredeskul, I. V. Shadrivov, and Yu. S. Kivshar, Phys. Rev. E \textbf{73}, 056605 (2006).

\bibitem{And} P. W. Anderson, Phys. Rev. \textbf{109}, 1492 (1958).

\bibitem{john84} S. John, Phys. Rev. Lett., \textbf{53}, 2169 (1984).

\bibitem{LGP} I. M. Lifshits, S. A. Gredeskul, and L. A. Pastur, \textit{%
Introduction to the Theory of Disordered Systems} (Wiley, New York 1987).

\bibitem{Ping91} P. Sheng, \textit{Scattering and localization of classical
waves in random media } (Singapore: World Scientific 1991).

\bibitem{FG} V. D. Freilikher and S.A. Gredeskul, \textit{Progress in Optics}
\textbf{30}, 137 (1992).

\bibitem{GMP} S. A. Gredeskul, A. V. Marchenko, and L. A. Pastur, \textit{%
Surveys in Applied Mathematics}, v. 2 (Plenum Press, New York 1995), p. 63.

\bibitem{Sheng06} P. Sheng, \textit{Introduction to Wave Scattering,
Localization, and Mesoscopic Phenomena} (Springer-Verlag, Heidelberg, 2006).

\bibitem{Mott} N. Mott and Twose, Adv. Phys. \textbf{10}, 107 (1961).

\bibitem{Israilev} F. M. Israilev, N. M. Makarov, E. J. Torres-Herrera,
arXiv:0911.0966v1 [cond-mat.mes-hall], 5 Nov. 2009.

\bibitem{Lyapunov} Y. Dong and X. Zhang, Phys. Lett. A \textbf{359}, 542
(2006).

\bibitem{we} A. A. Asatryan, L. C. Botten, M. A. Byrne, V. D. Freilikher, S. A. Gredeskul, I. V. Shadrivov, R. C. McPhedran, and Yu. S. Kivshar, Phys. Rev. Lett. \textbf{99}, 193902 (2007).

\bibitem{Baluni} V. Baluni and J. Willemsen, Phys. Rev. A \textbf{31}, 3358
(1985).

\bibitem{random} E. M. Nascimento, F. A. B. F. de Moura, and M. L. Lyra, Optics Express \textbf{16}, 6860 (2008).

\bibitem{single} P. Han, C. T. Chan, and Z. Q. Zhang, Phys. Rev. B \textbf{77}%
, 115332 (2008).

\bibitem{deSterke} C. Martijn de Sterke, and R. C. McPhedran, Phys. Rev. B \textbf{%
47}, 7780 (1993).

\bibitem{Asatrian2} A. A. Asatryan, N. A. Nicorovici, L. C. Botten, C. M. de
Sterke, P. A. Robinson, and R. C. McPhedran, Phys. Rev. B \textbf{57}, 13535
(1998).

\bibitem{Luna} G. A. Luna-Acosta, F. M. Izrailev, N. M. Makarov, U. Kuhl, and
H.-J. St$\ddot{o}$ckmann, Phys. Rev. B \textbf{80}, 115112 (2009).

\bibitem{Rytov} S. M. Rytov, Yu. A. Kravtsov, and V. I. Tatarskii, \textit{%
Principles of Statistical Radiophysics} (Springer, Berlin, 1987).

\bibitem{Liansky-1995} V.D. Freilikher, B.A. Liansky, I.V. Yurkevich, A.A.
Maradudin, A.R. McGurn, Phys. Rev. E {\bf 51}, 6301 (1995).

\bibitem{Asatryan99} A. A. Asatryan, P. A. Robinson, L. C. Botten, R. C. McPhedran, N. A. Nicorovici and C. Martijn de Sterke, Phys. Rev. E {\bf 60}, 6118 (1999).

\bibitem{Altshu} B. L. Altshuler, A. G. Aronov, and B. Z. Spivak, JETP
Lett. {\bf 32}, 94 (1981).

\bibitem{Lifshits} I. M. Lifshitz and V. Ya. Kirpichenkov, Sov. Phys. JETP
\textbf{50}, 499 (1979) [Zh. Eksp. Teor. Fiz. \textbf{77}, 989 (1979)].

\bibitem{Azbel} M. Ya. Azbel and P. Soven, Phys. Rev. B \textbf{27}, 831
(1983).

\bibitem{Bliokh-2} K. Yu. Bliokh, Yu. P. Bliokh, V. Freilikher, S. Savel\"{\i}ev, and F. Nori, Rev. Mod. Phys. \textbf{80}, 1201 (2008).

\bibitem{FrePusYur} V. Freilikher, M. Pustilnik, and I. Yurkevich, Phys.
Rev. B \textbf{50}, 6017 (1994).



\end{thebibliography}
\end{document}